\documentclass[prd,aps,nofootinbib,floatfix,11pt]{revtex4}

\usepackage{amsmath,graphicx,epsfig,amssymb,dsfont,mathtools}
\usepackage[usenames]{color}
\usepackage{ulem} 
\usepackage{bigstrut}
\usepackage{slashed}
\usepackage{multirow}
\usepackage{subfigure}

\allowdisplaybreaks

\begin{document}
	
	\title{Weak decays of doubly heavy baryons: the $1/2\to3/2$ case}
	
	\author{Zhen-Xing Zhao$^{1}$~\footnote{Email:star\_0027@sjtu.edu.cn}}
	
	\affiliation{$^{1}$ INPAC, Shanghai Key Laboratory for Particle Physics and Cosmology,
		\\
		MOE Key Laboratory for Particle Physics, Astrophysics and Cosmology,
		\\
		School of Physics and Astronomy, Shanghai Jiao-Tong University, Shanghai
		200240, P.R. China }
		
\begin{abstract}
As a continuation of our previous works, we investigate the weak decays of doubly heavy baryons into a spin-$3/2$ singly  or doubly heavy baryon.  Light-front
	approach  is  adopted to handle the dynamics in  
	the transitions,  in which the two spectator quarks are approximated as a diquark. Results for form factors  are then used  to calculate  
	decay widths of semi-leptonic and nonleptonic processes. The flavor  SU(3) symmetry and  symmetry breaking effects   in  semi-leptonic
	decays modes are explored, and we point out that in charm sector, there are sizable symmetry breaking effects.   For   nonleptonic decay modes, we study only the factorizable channels induced by the external W-emission. We find that
 branching fractions  for most 1/2 to 3/2 transitions are approximately one order of magnitude smaller
	than the corresponding ones for the 1/2 to 1/2  transitions.  Parametric uncertainties   
	are also investigated in detail. This work, together with our previous works,
	are beneficial  to  the  experimental studies   of doubly heavy baryons at
	LHC and other experiments.
\end{abstract}
	\maketitle
	
\section{Introduction}

Quite recently,   LHCb collaboration reported the discovery of a doubly
charmed baryon $\Xi_{cc}^{++}$ with the mass given as~\cite{Aaij:2017ueg}
\begin{equation}
	m_{\Xi_{cc}^{++}}=(3621.40\pm0.72\pm0.27\pm0.14){\rm MeV}.\label{eq:LHCb_measurement}
\end{equation}
This discovery has already triggered great theoretical interests on the study of  doubly heavy
baryons from various aspects~\cite{Chen:2017sbg,Yu:2017zst,Wang:2017mqp,Li:2017cfz,Meng:2017udf,Wang:2017azm,Karliner:2017qjm,Gutsche:2017hux,Li:2017pxa,Guo:2017vcf,Lu:2017meb,Xiao:2017udy,Sharma:2017txj,Ma:2017nik,Meng:2017dni,Li:2017ndo,Wang:2017qvg,Shi:2017dto,Hu:2017dzi,1803.01476}.  Also after observing the first decay mode, $\Xi_{cc}^{++}\to\Lambda_{c}^{+}K^{-}\pi^{+}\pi^{+}$,  LHCb collaboration is continuing the  experimental analyses  of doubly heavy
baryon decays. This includes, but is not limited to,
searches for the  $\Xi_{cc}^{+}$ and   charmed-beauty
$\Xi_{bc}$ baryons~\cite{Traill:2017zbs}.  Comprehensive theoretical studies on weak decays must be performed and  the golden modes for discoveries  must be derived in order to optimize the experimental resources.  In our previous work \cite{Wang:2017mqp}, we have
presented the calculation of 1/2 to 1/2 weak decays. It is generally anticipated 
that the 1/2 to 3/2 processes will also be important. For instance, it is very likely that  the $\Xi_{cc}^{++}\to\Lambda_{c}^{+}K^{-}\pi^{+}\pi^{+}$ comes from more than one  intermediate $1/2\to 3/2$ transitions.

A doubly heavy baryon is composed of two heavy
quarks and one light quark. Light flavor SU(3) symmetry arranges the
doubly heavy baryons into the presentation $\boldsymbol{3}$. For
spin-1/2 doubly heavy baryons, we have $\Xi_{cc}^{++,+}$ and $\Omega_{cc}^{+}$
in the $cc$ sector, $\Xi_{bb}^{0,-}$ and $\Omega_{bb}^{-}$ in the $bb$ sector. There are two sets of baryons for $bc$ sector depending on the
symmetric property under interchange of $b$ and $c$ quarks. If it is symmetric
under interchange of $b$ and $c$ quarks, this set is denoted by $\Xi_{bc}^{+,0}$and
$\Omega_{bc}^{0}$, while for the asymmetric case, the corresponding
set is denoted by $\Xi_{bc}^{\prime+,\prime0}$ and $\Omega_{bc}^{\prime0}$.\footnote{It should be noted that the convention here for $bc$ sector is the
opposite of that in Ref. \cite{Brown:2014ena}.} In reality these two sets probably
mix with each other, which is not taken into account   in
this work. Spin-3/2 doubly heavy baryons have the same flavor wave
functions with but different spin structures compared to the spin-1/2
counterparts. The quantum numbers  of low-lying doubly heavy baryons can be
found in Table~\ref{Tab:JPC}.

\begin{table*}[!htb]
	{\footnotesize{}\caption{Quantum numbers and quark content for the ground state of doubly heavy
			baryons. The $S_{h}^{\pi}$ denotes the spin of the heavy quark system.
			The light quark $q$ corresponds to $u,d$ quark. }
		\label{Tab:JPC} }{\footnotesize \par}
	\centering{}{\footnotesize{}}%
	\begin{tabular}{cccc|cccc}
		\hline 
		{\footnotesize{}Baryon } & {\footnotesize{}Quark Content } & {\footnotesize{}$S_{h}^{\pi}$ } & {\footnotesize{}$J^{P}$ } & {\footnotesize{}Baryon } & {\footnotesize{}Quark Content } & {\footnotesize{}$S_{h}^{\pi}$ } & {\footnotesize{}$J^{P}$ }\tabularnewline
		\hline 
		{\footnotesize{}$\Xi_{cc}$ } & {\footnotesize{}$\{cc\}q$ } & {\footnotesize{}$1^{+}$ } & {\footnotesize{}$1/2^{+}$ } & {\footnotesize{}$\Xi_{bb}$ } & {\footnotesize{}$\{bb\}q$ } & {\footnotesize{}$1^{+}$ } & {\footnotesize{}$1/2^{+}$ }\tabularnewline
		{\footnotesize{}$\Xi_{cc}^{*}$ } & {\footnotesize{}$\{cc\}q$ } & {\footnotesize{}$1^{+}$ } & {\footnotesize{}$3/2^{+}$ } & {\footnotesize{}$\Xi_{bb}^{*}$ } & {\footnotesize{}$\{bb\}q$ } & {\footnotesize{}$1^{+}$ } & {\footnotesize{}$3/2^{+}$ }\tabularnewline
		\hline 
		{\footnotesize{}$\Omega_{cc}$ } & {\footnotesize{}$\{cc\}s$ } & {\footnotesize{}$1^{+}$ } & {\footnotesize{}$1/2^{+}$ } & {\footnotesize{}$\Omega_{bb}$ } & {\footnotesize{}$\{bb\}s$ } & {\footnotesize{}$1^{+}$ } & {\footnotesize{}$1/2^{+}$ }\tabularnewline
		{\footnotesize{}$\Omega_{cc}^{*}$ } & {\footnotesize{}$\{cc\}s$ } & {\footnotesize{}$1^{+}$ } & {\footnotesize{}$3/2^{+}$ } & {\footnotesize{}$\Omega_{bb}^{*}$ } & {\footnotesize{}$\{bb\}s$ } & {\footnotesize{}$1^{+}$ } & {\footnotesize{}$3/2^{+}$ }\tabularnewline
		\hline 
		{\footnotesize{}$\Xi_{bc}^{\prime}$ } & {\footnotesize{}$[bc]q$ } & {\footnotesize{}$0^{+}$ } & {\footnotesize{}$1/2^{+}$ } & {\footnotesize{}$\Omega_{bc}^{\prime}$ } & {\footnotesize{}$[bc]s$ } & {\footnotesize{}$0^{+}$ } & {\footnotesize{}$1/2^{+}$ }\tabularnewline
		{\footnotesize{}$\Xi_{bc}$ } & {\footnotesize{}$\{bc\}q$ } & {\footnotesize{}$1^{+}$ } & {\footnotesize{}$1/2^{+}$ } & {\footnotesize{}$\Omega_{bc}$ } & {\footnotesize{}$\{bc\}s$ } & {\footnotesize{}$1^{+}$ } & {\footnotesize{}$1/2^{+}$ }\tabularnewline
		{\footnotesize{}$\Xi_{bc}^{*}$ } & {\footnotesize{}$\{bc\}q$ } & {\footnotesize{}$1^{+}$ } & {\footnotesize{}$3/2^{+}$ } & {\footnotesize{}$\Omega_{bc}^{*}$ } & {\footnotesize{}$\{bc\}s$ } & {\footnotesize{}$1^{+}$ } & {\footnotesize{}$3/2^{+}$ }\tabularnewline
		\hline 
	\end{tabular}{\footnotesize \par}
\end{table*}

\begin{figure}[!]
	\includegraphics[width=0.5\columnwidth]{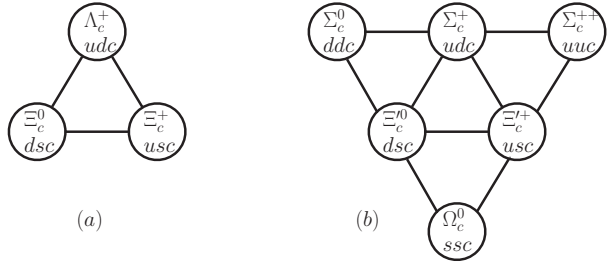} \caption{Anti-triplets (panel a) and sextets (panel b) of charmed baryons with one charm quark and two light quarks. These baryons are spin-$1/2$, while spin-$3/2$ baryons constitute another sextets. }
	\label{fig:singly_heavy} 
\end{figure}

The decay final state of the $\Xi_{cc}$ and $\Omega_{cc}$ contains
  baryons with one charm quark and two light quarks. Light flavor
SU(3) symmetry arranges them into the presentations $\boldsymbol{3}\otimes\boldsymbol{3}=\boldsymbol{6}\oplus\bar{\boldsymbol{3}}$,
as can be seen from Fig.~\ref{fig:singly_heavy}. The irreducible representation $\bar{\boldsymbol{3}}$ is composed
of $\Lambda_{c}^{+}$ and $\Xi_{c}^{+,0}$ while the sextet is composed
of $\Sigma_{c}^{++,+,0}$, $\Xi_{c}^{\prime+,\prime0}$ and $\Omega_{c}^{0}$.
They all have spin 1/2, while for spin-3/2 sextet, we will denote
them by $\Sigma_{c}^{*++,*+,*0}$, $\Xi_{c}^{\prime*+,\prime*0}$
and $\Omega_{c}^{*0}$. The singly bottom baryons can be analyzed
in a similar way.

To be explicit, we will investigate the following decay modes of doubly
heavy baryons. 
\begin{itemize}
	\item $cc$ sector 
	\begin{align*}
		\Xi_{cc}^{++}(ccu) & \to\Sigma_{c}^{*+}(dcu)/\Xi_{c}^{\prime*+}(scu),\\
		\Xi_{cc}^{+}(ccd) & \to\Sigma_{c}^{*0}(dcd)/\Xi_{c}^{\prime*0}(scd),\\
		\Omega_{cc}^{+}(ccs) & \to\Xi_{c}^{\prime*0}(dcs)/\Omega_{c}^{*0}(scs),
	\end{align*}
	\item $bb$ sector 
	\begin{align*}
		\Xi_{bb}^{0}(bbu) & \to\Sigma_{b}^{*+}(ubu)/\Xi_{bc}^{*+}(cbu),\\
		\Xi_{bb}^{-}(bbd) & \to\Sigma_{b}^{*0}(ubd)/\Xi_{bc}^{*0}(cbd),\\
		\Omega_{bb}^{-}(bbs) & \to\Xi_{b}^{\prime*0}(ubs)/\Omega_{bc}^{*0}(cbs),
	\end{align*}
	\item $bc$ sector with the $c$ quark decay 
	\begin{align*}
		\Xi_{bc}^{+}(cbu)/\Xi_{bc}^{\prime+}(cbu) & \to\Sigma_{b}^{*0}(dbu)/\Xi_{b}^{\prime*0}(sbu),\\
		\Xi_{bc}^{0}(cbd)/\Xi_{bc}^{\prime0}(cbd) & \to\Sigma_{b}^{*-}(dbd)/\Xi_{b}^{\prime*-}(sbd),\\
		\Omega_{bc}^{0}(cbs)/\Omega_{bc}^{\prime0}(cbs) & \to\Xi_{b}^{\prime*-}(dbs)/\Omega_{b}^{*-}(sbs),
	\end{align*}
	\item $bc$ sector with the $b$ quark decay 
	\begin{align*}
		\Xi_{bc}^{+}(bcu)/\Xi_{bc}^{\prime+}(bcu) & \to\Sigma_{c}^{*++}(ucu)/\Xi_{cc}^{*++}(ccu),\\
		\Xi_{bc}^{0}(bcd)/\Xi_{bc}^{\prime0}(bcd) & \to\Sigma_{c}^{*+}(ucd)/\Xi_{cc}^{*+}(ccd),\\
		\Omega_{bc}^{0}(bcs)/\Omega_{bc}^{\prime0}(bcs) & \to\Xi_{c}^{\prime*+}(ucs)/\Omega_{cc}^{*+}(ccs).
	\end{align*}
\end{itemize}
In the above, the quark components have been explicitly shown in the
brackets, in which the quarks that participate in weak decay are placed  first.

\begin{figure}[!]
	\includegraphics[width=0.4\columnwidth]{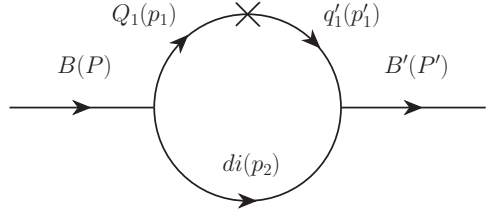}
	\caption{Feynman diagrams for baryon-baryon transitions in the diquark picture.
		$P^{(\prime)}$ is the momentum of the incoming (outgoing) baryon,
		$p_{1}^{(\prime)}$ is the initial (final) quark momentum, $p_{2}$
		is the diquark momentum and the cross mark denotes the  weak interaction.}
	\label{fig:decay} 
\end{figure}

To deal with the dynamics in the decay, we will
adopt the light front approach, which has been widely used to study the properties of mesons~\cite{Jaus:1999zv,Jaus:1989au,Jaus:1991cy,Cheng:1996if,Cheng:2003sm,Cheng:2004yj,Ke:2009ed,Ke:2009mn,Cheng:2009ms,Lu:2007sg,Wang:2007sxa,Wang:2008xt,Wang:2008ci,Wang:2009mi,Chen:2009qk,Li:2010bb,Verma:2011yw,Shi:2016gqt}.
Its application to the baryon sector can be found in Refs.~\cite{Ke:2007tg,Wei:2009np,Ke:2012wa,Zhu:2018jet,Ke:2017eqo}, in which the two spectator quarks are viewed as a diquark. In this scheme, the role of the diquark system  is similar to that of the antiquark in the meson case, as can be seen from Fig.~\ref{fig:decay}. Following the same method, in this work we will study the $1/2\to 3/2$ transition~\cite{Ke:2017eqo}, where the spectator is a $1^+$ diquark system. 

The authors of Ref.~\cite{Wang:2017azm} have investigated the doubly heavy baryon decays with the help of flavor SU(3) symmetry. Based on the available data, a great number of decay modes ranging from semi-leptonic decays to multi-body nonleptonic decays can be predicted. However, in the $c$ quark decay, SU(3) symmetry breaking effects may be sizable and can not be omitted. A quantitative study of SU(3) symmetry breaking effects will be conducted within the light-front approach.

The rest of the paper is arranged as follows. In Sec. II, we will
present briefly the framework of light-front approach under the diquark
picture, and flavor-spin wave functions will also be discussed. Numerical results are shown in Sec.~III, including the results for
form factors, predictions on semi-leptonic and nonleptonic decay widths,
detailed discussions on the SU(3) symmetry, the error estimates and a comparison with  the previous 1/2 to 1/2 results. A brief summary and
  discussions on   future improvements are given in the last
section.


\section{Theoretical framework}

Theoretical framework for $1/2 \to 3/2$ transition
will be briefly introduced in the first subsection, including the
definitions of the states for spin-1/2 and spin-3/2 baryons, and the extraction
of form factors. More details can be found in \cite{Ke:2017eqo,Ke:2007tg}.
Flavor-spin wave functions will be given in the second subsection.

\subsection{Light-front approach}

\label{subsec:light-front approach}

In the framework of light-front approach, the wave functions of $1/2^{+}$
baryon with an axial-vector diquark is expressed as
\begin{eqnarray}
	|{\cal B}(P,S=\frac{1}{2},S_{z})\rangle & = & \int\{d^{3}p_{1}\}\{d^{3}p_{2}\}2(2\pi)^{3}\delta^{3}(\tilde{P}-\tilde{p}_{1}-\tilde{p}_{2})\nonumber \\
	&  & \times\sum_{\lambda_{1}\lambda_{2}}\Psi^{SS_{z}}(\tilde{p}_{1},\tilde{p}_{2},\lambda_{1},\lambda_{2})|Q(p_{1},\lambda_{1})\{di\}(p_{2},\lambda_{2})\rangle.\label{eq:1/2-baryon}
\end{eqnarray}
Here
\begin{eqnarray}
	\Psi^{SS_{z}}(\tilde{p}_{1},\tilde{p}_{2},\lambda_{1},\lambda_{2}) & = & \frac{A}{\sqrt{2(p_{1}\cdot\bar{P}+m_{1}M_{0})}}\bar{u}(p_{1},\lambda_{1})\Gamma u(\bar{P},S_{z})\phi(x,k_{\perp})\label{eq:momentum_wave_fuction_1/2}
\end{eqnarray}
with
\begin{equation}
	\Gamma=-\frac{1}{\sqrt{3}}\gamma_{5}\slashed\epsilon^{*}(p_{2},\lambda_{2}),\quad m_{1}=m_{Q},\quad\bar{P}=p_{1}+p_{2},
\end{equation}
\begin{equation}
	\phi=4\left(\frac{\pi}{\beta^{2}}\right)^{3/4}\sqrt{\frac{e_{1}e_{2}}{x_{1}x_{2}M_{0}}}\exp\left(\frac{-\vec{k}^{2}}{2\beta^{2}}\right)\label{eq:Gauss}
\end{equation}
and 
\begin{equation}
	A=\sqrt{\frac{3(m_{1}M_{0}+p_{1}\cdot\bar{P})}{3m_{1}M_{0}+p_{1}\cdot\bar{P}+2(p_{1}\cdot p_{2})(p_{2}\cdot\bar{P})/m_{2}^{2}}.}
\end{equation}

In analog to the $1/2^{+}$ baryon case, $3/2^{+}$
baryon state has a similar expression like Eq. (\ref{eq:1/2-baryon})
but with Eq. (\ref{eq:momentum_wave_fuction_1/2})
being replaced by
\begin{equation}
	\Psi^{SS_{z}}(\tilde{p}_{1},\tilde{p}_{2},\lambda_{1},\lambda_{2})=\frac{A^{\prime}}{\sqrt{2(p_{1}\cdot\bar{P}+m_{1}M_{0})}}\bar{u}(p_{1},\lambda_{1})\epsilon^{\alpha*}(p_{2},\lambda_{2})u_{\alpha}(\bar{P},S_{z})\phi(x,k_{\perp}),\label{eq:momentum_wave_fuction_3/2}
\end{equation}
where 
\begin{equation}
	A^{\prime}=\sqrt{\frac{3m_{2}^{2}M_{0}^{2}}{2m_{2}^{2}M_{0}^{2}+(p_{2}\cdot\bar{P})^{2}}}.
\end{equation}

With the help of Eqs. (\ref{eq:1/2-baryon}), (\ref{eq:momentum_wave_fuction_1/2})
and (\ref{eq:momentum_wave_fuction_3/2}), the transition matrix element
can be derived as 
\begin{eqnarray}
	&  & \langle{\cal B}_{f}(P^{\prime},S^{\prime}=\frac{3}{2},S_{z}^{\prime})|\bar{q}\gamma^{\mu}(1-\gamma_{5})Q|{\cal B}_{i}(P,S=\frac{1}{2},S_{z})\rangle\nonumber \\
	& = & \int\{d^{3}p_{2}\}\frac{\varphi^{\prime}(x^{\prime},k_{\perp}^{\prime})\varphi(x,k_{\perp})}{2\sqrt{p_{1}^{+}p_{1}^{\prime+}(p_{1}\cdot\bar{P}+m_{1}M_{0})(p_{1}^{\prime}\cdot\bar{P}^{\prime}+m_{1}^{\prime}M_{0}^{\prime})}}\nonumber \\
	&  & \times\sum_{\lambda_{2}}\bar{u}_{\alpha}(\bar{P}^{\prime},S_{z}^{\prime})\left[\epsilon^{\alpha}(\lambda_{2})(\slashed p_{1}^{\prime}+m_{1}^{\prime})\gamma^{\mu}(1-\gamma_{5})(\slashed p_{1}+m_{1})\left(-\frac{1}{\sqrt{3}}\gamma_{5}\slashed\epsilon^{*}(\lambda_{2})\right)\right]u(\bar{P},S_{z}),\label{eq:matrix_element_1}
\end{eqnarray}
where 
\[
m_{1}=m_{Q},\quad m_{1}^{\prime}=m_{q},\quad m_{2}=m_{\{di\}},
\]
and $\varphi^{(\prime)}=A^{(\prime)}\phi^{(\prime)}$, $Q$ ($q$)
represents the quark $b/c$ ($u/d/s/c$) in the initial (final) state,
$p_{1}$ ($p_{1}^{\prime}$) denotes its four-momentum, $P$ ($P^{\prime}$)
stands for the four-momentum of ${\cal B}_{i}$ (${\cal B}_{f}$).
The form factors for $1/2\to3/2$ transition are parameterized as
\begin{eqnarray}
	\langle{\cal B}_{f}(P^{\prime},S^{\prime}=\frac{3}{2},S_{z}^{\prime})|\bar{q}\gamma^{\mu}Q|{\cal B}_{i}(P,S=\frac{1}{2},S_{z})\rangle & = & \bar{u}_{\alpha}(P^{\prime},S_{z}^{\prime})\Bigg[\gamma^{\mu}P^{\alpha}\frac{f_{1}(q^{2})}{M}+\frac{f_{2}(q^{2})}{M^{2}}P^{\alpha}P^{\mu}\nonumber \\
	&  & +\frac{f_{3}(q^{2})}{MM^{\prime}}P^{\alpha}P^{\prime\mu}+f_{4}(q^{2})g^{\alpha\mu}\Bigg]\gamma_{5}u(P,S_{z}),\label{eq:matrix_element_2V}\\
	\langle{\cal B}_{f}(P^{\prime},S^{\prime}=\frac{3}{2},S_{z}^{\prime})|\bar{q}\gamma^{\mu}\gamma_{5}Q|{\cal B}_{i}(P,S=\frac{1}{2},S_{z})\rangle & = & \bar{u}_{\alpha}(P^{\prime},S_{z}^{\prime})\Bigg[\gamma^{\mu}P^{\alpha}\frac{g_{1}(q^{2})}{M}+\frac{g_{2}(q^{2})}{M^{2}}P^{\alpha}P^{\mu}\nonumber \\
	&  & +\frac{g_{3}(q^{2})}{MM^{\prime}}P^{\alpha}P^{\prime\mu}+g_{4}(q^{2})g^{\alpha\mu}\Bigg]u(P,S_{z}).\label{eq:matrix_element_2A}
\end{eqnarray} 
Here $q=P-P^{\prime}$, and $f_{i}$, $g_{i}$ are the form factors.

These form factors $f_{i}$ and $g_{i}$ can be extracted in the following
way \cite{Ke:2017eqo}. Multiplying Eq. (\ref{eq:matrix_element_1}) by
$\bar{u}(P,S_{z})(\Gamma_{5}^{\mu\beta})_{i}u_{\beta}(P^{\prime},S_{z}^{\prime})$
with $(\Gamma_{5}^{\mu\beta})_{i}=\{\gamma^{\mu}P^{\beta},P^{\prime\mu}P^{\beta},P^{\mu}P^{\beta},g^{\mu\beta}\}\gamma_{5}$
respectively, and taking the approximation $P^{(\prime)}\to\bar{P}^{(\prime)}$
within the integral, and then summing over the polarizations in the
initial and final states, one can arrive at

\begin{eqnarray}
	F_{i} & = & \int\{d^{3}p_{2}\}\frac{\varphi^{\prime}(x^{\prime},k_{\perp}^{\prime})\varphi(x,k_{\perp})}{2\sqrt{p_{1}^{+}p_{1}^{\prime+}(p_{1}\cdot\bar{P}+m_{1}M_{0})(p_{1}^{\prime}\cdot\bar{P}^{\prime}+m_{1}^{\prime}M_{0}^{\prime})}}\sum_{S_{z}^{\prime}S_{z}\lambda_{2}}{\rm Tr}\Bigg\{ u_{\beta}(\bar{P}^{\prime},S_{z}^{\prime})\bar{u}_{\alpha}(\bar{P}^{\prime},S_{z}^{\prime})\nonumber \\
	&  & \times\epsilon^{\alpha}(\lambda_{2})(\slashed p_{1}^{\prime}+m_{1}^{\prime})\gamma_{\mu}(\slashed p_{1}+m_{1})\left(-\frac{1}{\sqrt{3}}\gamma_{5}\slashed\epsilon^{*}(\lambda_{2})\right)u(\bar{P},S_{z})\bar{u}(\bar{P},S_{z})(\bar{\Gamma}_{5}^{\mu\beta})_{i}\Bigg\}\label{eq:Fi_1}
\end{eqnarray}
with $(\bar{\Gamma}_{5}^{\mu\beta})_{i}=\{\gamma^{\mu}\bar{P}^{\beta},\bar{P}^{\prime\mu}\bar{P}^{\beta},\bar{P}^{\mu}\bar{P}^{\beta},g^{\mu\beta}\}\gamma_{5}$.

Multiplying the difference of Eq. (\ref{eq:matrix_element_2V}) and
Eq. (\ref{eq:matrix_element_2A}) by the same factor $\bar{u}(P,S_{z})(\Gamma_{5}^{\mu\beta})_{i}u_{\beta}(P^{\prime},S_{z}^{\prime})$,
and also summing over the polarizations in the initial and final states,
one can arrive at
\begin{eqnarray}
	F_{i} & = & {\rm Tr}\Bigg\{ u_{\beta}(P^{\prime},S_{z}^{\prime})\bar{u}_{\alpha}(P^{\prime},S_{z}^{\prime})\left[\gamma^{\mu}P^{\alpha}\frac{f_{1}(q^{2})}{M}+\frac{f_{2}(q^{2})}{M^{2}}P^{\alpha}P^{\mu}+\frac{f_{3}(q^{2})}{MM^{\prime}}P^{\alpha}P^{\prime\mu}+f_{4}(q^{2})g^{\alpha\mu}\right]\gamma_{5}\nonumber \\
	&  & \times u(P,S_{z})\bar{u}(P,S_{z})(\Gamma_{5}^{\mu\beta})_{i}\Bigg\}.\label{eq:Fi_2}
\end{eqnarray}

The form factors $f_{i}$ can then be extracted by equating Eqs. (\ref{eq:Fi_1})
and (\ref{eq:Fi_2}). With the same method, one can obtain the form
factors $g_{i}$.

\subsection{Flavor-spin wave functions}

In subsection \ref{subsec:light-front approach}, the  flavor-spin wave
function was not taken into account.  
We consider first  the initial state.
For the doubly charmed baryons, the wave functions are given as 
\begin{equation}
	\mathcal{B}_{cc}=\frac{1}{\sqrt{2}}\left[\left(-\frac{\sqrt{3}}{2}c^{1}(c^{2}q)_{S}+\frac{1}{2}c^{1}(c^{2}q)_{A}\right)+(c^{1}\leftrightarrow c^{2})\right],
\end{equation}
with $q=u$, $d$ or $s$ for $\Xi_{cc}^{++}$, $\Xi_{cc}^{+}$ or
$\Omega_{cc}^{+}$, respectively. It is similar for the doubly bottom baryons. For the bottom-charm baryons,
there are two sets of states, with $bc$ as a scalar or an axial-vector
diquark. The wave functions of bottom-charm baryons with an axial-vector
$bc$ diquark are 
\begin{align}
	\mathcal{B}_{bc} & =-\frac{\sqrt{3}}{2}b(cq)_{S}+\frac{1}{2}b(cq)_{A}=-\frac{\sqrt{3}}{2}c(bq)_{S}+\frac{1}{2}c(bq)_{A},\label{eq:flavor_spin_bc}
\end{align}
while those with a scalar $bc$ diquark are given as 
\begin{align}
	\mathcal{B}_{bc}^{\prime} & =-\frac{1}{2}b(cq)_{S}-\frac{\sqrt{3}}{2}b(cq)_{A}=\frac{1}{2}c(bq)_{S}+\frac{\sqrt{3}}{2}c(bq)_{A},\label{eq:flavor_spin_bcp}
\end{align}
with $q=u$, $d$ or $s$ for $\Xi_{bc}^{(\prime)+}$, $\Xi_{bc}^{(\prime)0}$
or $\Omega_{bc}^{(\prime)0}$, respectively. Note that the conventions 
for ${\cal B}_{bc}^{(\prime)}$ in Ref. \cite{Brown:2014ena} are opposite to ours.

For  the final state, the spin-3/2
baryon with quark contents of $Qqq^{\prime}$ has
\begin{equation}
{\cal B}_{Qqq^{\prime}}^{*}=q(Qq^{\prime})_{A}=q^{\prime}(Qq)_{A},
\end{equation}
while for the $Qqq$ baryon, an additional factor $\sqrt{2}$ should
be added. For the spin-3/2 baryon with quark contents of $QQ^{\prime}q$,
we have
\begin{equation}
{\cal B}_{QQ^{\prime}q}^{*}=Q(Q^{\prime}q)_{A}=Q^{\prime}(Qq)_{A},
\end{equation}
while for the $QQq$ baryon, an additional factor $\sqrt{2}$ should
be added. Here the asterisk denotes that the baryon is spin-3/2 and
$q^{(\prime)}=u,d,s$.

Finally, the overlapping factors are determined by taking the inner
product of the flavor-spin wave functions in the initial and final
states. The corresponding results are collected in Table \ref{Tab:overlapping_factors}.

\begin{table}
	\caption{Flavor-spin space overlapping  factors}
	{\footnotesize{}\label{Tab:overlapping_factors} }{\footnotesize \par}
	\begin{tabular}{c|c|c|c}
		\hline 
		transitions & overlapping factors & transitions & overlapping factors\tabularnewline
		\hline 
		$\Xi_{cc}^{++}(ccu)\to\Sigma_{c}^{*+}(dcu)/\Xi_{c}^{\prime*+}(scu),$ & $\frac{1}{\sqrt{2}},\quad\frac{1}{\sqrt{2}}$ & $\Xi_{bb}^{0}(bbu)\to\Sigma_{b}^{*+}(ubu)/\Xi_{bc}^{*+}(cbu),$ & $1,\quad\frac{1}{\sqrt{2}}$\tabularnewline
		\hline 
		$\Xi_{cc}^{+}(ccd)\to\Sigma_{c}^{*0}(dcd)/\Xi_{c}^{\prime*0}(scd),$ & $1,\quad\frac{1}{\sqrt{2}}$ & $\Xi_{bb}^{-}(bbd)\to\Sigma_{b}^{*0}(ubd)/\Xi_{bc}^{*0}(cbd),$ & $\frac{1}{\sqrt{2}},\quad\frac{1}{\sqrt{2}}$\tabularnewline
		\hline 
		$\Omega_{cc}^{+}(ccs)\to\Xi_{c}^{\prime*0}(dcs)/\Omega_{c}^{*0}(scs),$ & $\frac{1}{\sqrt{2}},\quad1$ & $\Omega_{bb}^{-}(bbs)\to\Xi_{b}^{\prime*0}(ubs)/\Omega_{bc}^{*0}(cbs),$ & $\frac{1}{\sqrt{2}},\quad\frac{1}{\sqrt{2}}$\tabularnewline
		\hline 
		$\Xi_{bc}^{+}(cbu)\to\Sigma_{b}^{*0}(dbu)/\Xi_{b}^{\prime*0}(sbu)$ & $\frac{1}{2},\quad\frac{1}{2}$ & $\Xi_{bc}^{+}(bcu)\to\Sigma_{c}^{*++}(ucu)/\Xi_{cc}^{*++}(ccu)$ & $\frac{\sqrt{2}}{2},\quad\frac{\sqrt{2}}{2}$\tabularnewline
		\hline 
		$\Xi_{bc}^{0}(cbd)\to\Sigma_{b}^{*-}(dbd)/\Xi_{b}^{\prime*-}(sbd)$ & $\frac{\sqrt{2}}{2},\quad\frac{1}{2}$ & $\Xi_{bc}^{0}(bcd)\to\Sigma_{c}^{*+}(ucd)/\Xi_{cc}^{*+}(ccd)$ & $\frac{1}{2},\quad\frac{\sqrt{2}}{2}$\tabularnewline
		\hline 
		$\Omega_{bc}^{0}(cbs)\to\Xi_{b}^{\prime*-}(dbs)/\Omega_{b}^{*-}(sbs)$ & $\frac{1}{2},\quad\frac{\sqrt{2}}{2}$ & $\Omega_{bc}^{0}(bcs)\to\Xi_{c}^{\prime*+}(ucs)/\Omega_{cc}^{*+}(ccs)$ & $\frac{1}{2},\quad\frac{\sqrt{2}}{2}$\tabularnewline
		\hline 
		$\Xi_{bc}^{\prime+}(cbu)\to\Sigma_{b}^{*0}(dbu)/\Xi_{b}^{\prime*0}(sbu)$ & $\frac{\sqrt{3}}{2},\quad\frac{\sqrt{3}}{2}$ & $\Xi_{bc}^{\prime+}(bcu)\to\Sigma_{c}^{*++}(ucu)/\Xi_{cc}^{*++}(ccu)$ & $-\frac{\sqrt{6}}{2},\quad-\frac{\sqrt{6}}{2}$\tabularnewline
		\hline 
		$\Xi_{bc}^{\prime0}(cbd)\to\Sigma_{b}^{*-}(dbd)/\Xi_{b}^{\prime*-}(sbd)$ & $\frac{\sqrt{6}}{2},\quad\frac{\sqrt{3}}{2}$ & $\Xi_{bc}^{\prime0}(bcd)\to\Sigma_{c}^{*+}(ucd)/\Xi_{cc}^{*+}(ccd)$ & $-\frac{\sqrt{3}}{2},\quad-\frac{\sqrt{6}}{2}$\tabularnewline
		\hline 
		$\Omega_{bc}^{\prime0}(cbs)\to\Xi_{b}^{\prime*-}(dbs)/\Omega_{b}^{*-}(sbs)$ & $\frac{\sqrt{3}}{2},\quad\frac{\sqrt{6}}{2}$ & $\Omega_{bc}^{\prime0}(bcs)\to\Xi_{c}^{\prime*+}(ucs)/\Omega_{cc}^{*+}(ccs)$ & $-\frac{\sqrt{3}}{2},\quad-\frac{\sqrt{6}}{2}$\tabularnewline
		\hline 
	\end{tabular}
\end{table}

\section{Numerical results and discussions}

All the inputs will be given in the first subsection. Numerical results
for form factors, semi-leptonic and nonleptonic decays will be shown subsequently. They are presented in this order: $cc$
sector, $bb$ sector, $bc$ sector with the $c$ quark decay, $bc$
sector with the $b$ quark decay, $bc^{\prime}$ sector with the $c$
quark decay, $bc^{\prime}$ sector with the $b$ quark decay. Some
discussions will also be given.

The variable $\omega$ will be introduced
\begin{equation}
	\omega\equiv v\cdot v^{\prime}=\frac{P\cdot P^{\prime}}{MM^{\prime}},\label{eq:omega}
\end{equation}
which can be easily changed to the squared momentum transfer $q^{2}$,
and vice versa.

The vectorial spinor for spin-$3/2$ baryon is given as
\begin{equation}
	u^{\alpha}=(\epsilon^{\alpha}-\frac{1}{3}(\gamma^{\alpha}+v^{\alpha})\slashed\epsilon)u
\end{equation}
with $v^{\alpha}=p^{\alpha}/m$, while its helicity eigenstate can
be found in Eq. (20) of Ref. \cite{Auvil:1966eao}
\begin{equation}
	u^{\alpha}(p,\lambda)=\sum_{\lambda_{1},\lambda_{2}}\langle\frac{1}{2},\lambda_{1},1,\lambda_{2}|\frac{3}{2},\lambda\rangle\times u(p,\lambda_{1})\epsilon^{\alpha}(p,\lambda_{2}).
\end{equation}

\subsection{Inputs}

The constituent  quark masses are given as (in units of GeV)~\cite{Lu:2007sg,Wang:2007sxa,Wang:2008xt,Wang:2008ci,Wang:2009mi,Chen:2009qk,Li:2010bb,Verma:2011yw,Shi:2016gqt} 
\begin{equation}
	m_{u}=m_{d}=0.25,\quad m_{s}=0.37,\quad m_{c}=1.4,\quad m_{b}=4.8.
\end{equation}
The masses of the
axial-vector diquarks are approximated by $m_{\{Qq\}}=m_{Q}+m_{q}$.
The shape parameters $\beta$ in Eq. (\ref{eq:Gauss}) are given as
(in units of GeV) \cite{Cheng:2003sm}
\begin{eqnarray}
	&  & \beta_{u\{cq\}}=\beta_{d\{cq\}}=0.470,\quad\beta_{s\{cq\}}=0.535,\quad\beta_{c\{cq\}}=0.753,\quad\beta_{b\{cq\}}=0.886,\nonumber \\
	&  & \beta_{u\{bq\}}=\beta_{d\{bq\}}=0.562,\quad\beta_{s\{bq\}}=0.623,\quad\beta_{c\{bq\}}=0.886,\quad\beta_{b\{bq\}}=1.472,
\end{eqnarray}
where $q=u,d,s$.

The masses and lifetimes of the parent baryons are collected in Table
\ref{Tab:parent_baryons}~\cite{Aaij:2017ueg,Brown:2014ena,Yu:2017zst,Guberina:1999qp,Karliner:2014gca,Kiselev:2001fw}. Note that, in
the Table \ref{Tab:parent_baryons}, the masses
and lifetimes of ${\cal B}_{bc}$ and ${\cal B}_{bc}^{\prime}$
are taken the same. Also note that we have taken a new value for the lifetime
of $\Omega_{cc}^{+}$ compared with our previous work \cite{Wang:2017mqp}. Because
according to Ref. \cite{Guberina:1999qp}, lifetimes of doubly charmed baryons should satisfy
the following pattern:
\begin{equation}
	\tau(\Xi_{cc}^{+})\sim\tau(\Omega_{cc}^{+})\ll\tau(\Xi_{cc}^{++}).
\end{equation}

\begin{table}[!htb]
	\caption{Masses (in units of GeV) and lifetimes (in units of fs) of doubly
		heavy baryons.}
	\label{Tab:parent_baryons} %
	\begin{tabular}{c|c|c|c|c|c|c|c|c|c}
		\hline 
		baryons  & $\Xi_{cc}^{++}$  & $\Xi_{cc}^{+}$  & $\Omega_{cc}^{+}$  & $\Xi_{bc}^{(\prime)+}$  & $\Xi_{bc}^{(\prime)0}$  & $\Omega_{bc}^{(\prime)0}$  & $\Xi_{bb}^{0}$  & $\Xi_{bb}^{-}$  & $\Omega_{bb}^{-}$ \tabularnewline
		\hline 
		masses  & $3.621$ \cite{Aaij:2017ueg}  & $3.621$ \cite{Aaij:2017ueg}  & $3.738$ \cite{Brown:2014ena}  & $6.943$ \cite{Brown:2014ena}  & $6.943$ \cite{Brown:2014ena}  & $6.998$ \cite{Brown:2014ena}  & $10.143$\cite{Brown:2014ena}  & $10.143$ \cite{Brown:2014ena}  & $10.273$\cite{Brown:2014ena}\tabularnewline
		\hline 
		lifetimes  & $300$~\cite{Yu:2017zst}  & $100$~\cite{Yu:2017zst}  & $100$~\cite{Guberina:1999qp} & $244$ \cite{Karliner:2014gca}  & $93$ \cite{Karliner:2014gca}  & $220$ \cite{Kiselev:2001fw}  & $370$ \cite{Karliner:2014gca}  & $370$ \cite{Karliner:2014gca}  & $800$\cite{Kiselev:2001fw}\tabularnewline
		\hline 
	\end{tabular}
\end{table}

The masses of  the final state baryons are given in Table \ref{Tab:daughter_baryons}
\cite{Olive:2016xmw,Brown:2014ena}. Fermi constant and CKM matrix elements are give as \cite{Olive:2016xmw}
\begin{align}
& G_{F}=1.166\times10^{-5}\ {\rm GeV}^{-2},\nonumber \\
& |V_{ud}|=0.974,\quad|V_{us}|=0.225,\quad|V_{ub}|=0.00357,\nonumber \\
& |V_{cd}|=0.225,\quad|V_{cs}|=0.974,\quad|V_{cb}|=0.0411.\label{eq:GFCKM}
\end{align}

\begin{table}[!]
	\caption{The masses of baryons in the final states \cite{Olive:2016xmw,Brown:2014ena}. }
	\label{Tab:daughter_baryons}
	\begin{tabular}{c|c|c|c|c|c|c|c|c}
		\hline 
		$\Sigma_{c}^{*++}$ & $\Sigma_{c}^{*+}$  & $\Sigma_{c}^{*0}$ & $\Xi_{c}^{\prime*+}$ & $\Xi_{c}^{\prime*0}$  & $\Omega_{c}^{*0}$ & $\Xi_{cc}^{*++}$ & $\Xi_{cc}^{*+}$  & $\Omega_{cc}^{*+}$\tabularnewline
		\hline 
		$2.518$  & $2.518$  & $2.518$  & $2.646$  & $2.646$  & $2.766$  & $3.692$  & $3.692$  & $3.822$ \tabularnewline
		\hline 
		$\Sigma_{b}^{*+}$ & $\Sigma_{b}^{*0}$  & $\Sigma_{b}^{*-}$ & $\Xi_{b}^{\prime*0}$ & $\Xi_{b}^{\prime*-}$  & $\Omega_{b}^{*-}$ & $\Xi_{bc}^{*+}$ & $\Xi_{bc}^{*0}$  & $\Omega_{bc}^{*0}$\tabularnewline
		\hline 
		$5.832$  & $5.833$  & $5.835$  & $5.949$  & $5.955$  & $6.085$  & $6.985$  & $6.985$  & $7.059$ \tabularnewline
		\hline 
	\end{tabular}
\end{table}

In the calculation of nonleptonic decays, these mesons will
present in the final states: $\pi,\rho,a_{1},K,K^{*},D,D^{*},D_{s},D_{s}^{*}$.
Their masses can be found in Ref.~\cite{Olive:2016xmw}, while their decay constants
are given as follows \cite{Cheng:2003sm,Shi:2016gqt,Carrasco:2014poa}:
\begin{align}
	f_{\pi} & =130.4{\rm MeV},\quad f_{\rho}=216{\rm MeV},\quad f_{a_{1}}=238{\rm MeV},\quad f_{K}=160{\rm MeV},\quad f_{K^{*}}=210{\rm MeV},\nonumber \\
	f_{D} & =207.4{\rm MeV},\quad f_{D^{*}}=220{\rm MeV},\quad f_{D_{s}}=247.2{\rm MeV},\quad f_{D_{s}^{*}}=247.2{\rm MeV}.
\end{align}
Wilson coefficients $a_{1}=C_{1}(\mu_{c})+C_{2}(\mu_{c})/3=1.07$~\cite{Li:2012cfa},
will   be used.

\subsection{Results for form factors}

To access the $q^2$-distribution, 
the following single pole structure is assumed for form factors:
\begin{equation}
	F(q^{2})=\frac{F(0)}{1-\frac{q^{2}}{m_{{\rm pole}^{2}}}},\label{eq:single_pole}
\end{equation}
$F(0)$ is the value of the form factors at $q^{2}=0$, the corresponding numerical
results predicted by the light-front approach are collected in Tables
\ref{Tab:ff_cc} to \ref{Tab:ff_bc_b}. For $c\to d/s$ decays, $m_{{\rm pole}}$
is taken as $1.87$ GeV, while for $b\to u/c$ decays, $m_{{\rm pole}}$
is taken as $5.28$ GeV and $6.28$ GeV, respectively. In practice, these quantities
are taken as the masses of $D$, $B$ and $B_{c}$ mesons. The discussion
for the validity of this assumption can be found in our previous work \cite{Shi:2016gqt}.

The physical form factor can be obtained by multiplying Eq. (\ref{eq:single_pole})
by the corresponding overlapping factor.

\begin{table}
	\caption{Values of form factors at $q^{2}=0$ for $cc$ sector. Single pole
		assumption in Eq. (\ref{eq:single_pole}) will be adopted, and $m_{{\rm pole}}$ is taken as $1.87\ {\rm GeV}$.}
	\label{Tab:ff_cc}
	
	\begin{tabular}{c|r|c|r}
		\hline 
		$F$  & $F(0)$  & $F$  & $F(0)$ \tabularnewline
		\hline 
		$f_{1}^{\Xi_{cc}\to\Sigma_{c}^{*}}$  & $-1.121$  & $g_{1}^{\Xi_{cc}\to\Sigma_{c}^{*}}$  & $-8.292$ \tabularnewline
		$f_{2}^{\Xi_{cc}\to\Sigma_{c}^{*}}$  & $1.764$  & $g_{2}^{\Xi_{cc}\to\Sigma_{c}^{*}}$  & $-0.156$ \tabularnewline
		$f_{3}^{\Xi_{cc}\to\Sigma_{c}^{*}}$  & $-3.793$  & $g_{3}^{\Xi_{cc}\to\Sigma_{c}^{*}}$  & $7.427$ \tabularnewline
		$f_{4}^{\Xi_{cc}\to\Sigma_{c}^{*}}$  & $-1.827$  & $g_{4}^{\Xi_{cc}\to\Sigma_{c}^{*}}$  & $0.295$ \tabularnewline
		\hline 
		$f_{1}^{\Xi_{cc}\to\Xi_{c}^{\prime*}}$  & $-1.318$  & $g_{1}^{\Xi_{cc}\to\Xi_{c}^{\prime*}}$  & $-14.180$ \tabularnewline
		$f_{2}^{\Xi_{cc}\to\Xi_{c}^{\prime*}}$  & $1.494$  & $g_{2}^{\Xi_{cc}\to\Xi_{c}^{\prime*}}$  & $-0.882$ \tabularnewline
		$f_{3}^{\Xi_{cc}\to\Xi_{c}^{\prime*}}$  & $-5.251$  & $g_{3}^{\Xi_{cc}\to\Xi_{c}^{\prime*}}$  & $13.600$ \tabularnewline
		$f_{4}^{\Xi_{cc}\to\Xi_{c}^{\prime*}}$  & $-2.147$  & $g_{4}^{\Xi_{cc}\to\Xi_{c}^{\prime*}}$  & $0.294$ \tabularnewline
		\hline 
		$f_{1}^{\Omega_{cc}^{+}\to\Xi_{c}^{\prime*0}}$  & $-1.154$  & $g_{1}^{\Omega_{cc}^{+}\to\Xi_{c}^{\prime*0}}$  & $-8.801$ \tabularnewline
		$f_{2}^{\Omega_{cc}^{+}\to\Xi_{c}^{\prime*0}}$  & $2.227$  & $g_{2}^{\Omega_{cc}^{+}\to\Xi_{c}^{\prime*0}}$  & $-0.118$ \tabularnewline
		$f_{3}^{\Omega_{cc}^{+}\to\Xi_{c}^{\prime*0}}$  & $-4.244$  & $g_{3}^{\Omega_{cc}^{+}\to\Xi_{c}^{\prime*0}}$  & $7.915$ \tabularnewline
		$f_{4}^{\Omega_{cc}^{+}\to\Xi_{c}^{\prime*0}}$  & $-1.896$  & $g_{4}^{\Omega_{cc}^{+}\to\Xi_{c}^{\prime*0}}$  & $0.298$ \tabularnewline
		\hline 
		$f_{1}^{\Omega_{cc}^{+}\to\Omega_{c}^{*0}}$  & $-1.339$  & $g_{1}^{\Omega_{cc}^{+}\to\Omega_{c}^{*0}}$  & $-14.470$ \tabularnewline
		$f_{2}^{\Omega_{cc}^{+}\to\Omega_{c}^{*0}}$  & $1.939$  & $g_{2}^{\Omega_{cc}^{+}\to\Omega_{c}^{*0}}$  & $-0.811$ \tabularnewline
		$f_{3}^{\Omega_{cc}^{+}\to\Omega_{c}^{*0}}$  & $-5.575$  & $g_{3}^{\Omega_{cc}^{+}\to\Omega_{c}^{*0}}$  & $13.850$ \tabularnewline
		$f_{4}^{\Omega_{cc}^{+}\to\Omega_{c}^{*0}}$  & $-2.204$  & $g_{4}^{\Omega_{cc}^{+}\to\Omega_{c}^{*0}}$  & $0.314$ \tabularnewline
		\hline 
	\end{tabular}
\end{table}

\begin{table}
	\caption{Values of form factors at $q^{2}=0$ for $bb$ sector. Single pole
		assumption in Eq. (\ref{eq:single_pole}) will be adopted, and $m_{{\rm pole}}$ is taken as follows: for $b\to q$ process, $m_{{\rm pole}}=5.28\ {\rm GeV}$
		while for $b\to c$ process, $m_{{\rm pole}}=6.28\ {\rm GeV}$.}
	\label{Tab:ff_bb}%
	\begin{tabular}{c|r|c|r}
		\hline 
		$F$  & $F(0)$  & $F$  & $F(0)$ \tabularnewline
		\hline 
		$f_{1}^{\Xi_{bb}\to\Sigma_{b}^{*}}$  & $-0.183$  & $g_{1}^{\Xi_{bb}\to\Sigma_{b}^{*}}$  & $-0.219$ \tabularnewline
		$f_{2}^{\Xi_{bb}\to\Sigma_{b}^{*}}$  & $0.230$  & $g_{2}^{\Xi_{bb}\to\Sigma_{b}^{*}}$  & $0.089$ \tabularnewline
		$f_{3}^{\Xi_{bb}\to\Sigma_{b}^{*}}$  & $-0.153$  & $g_{3}^{\Xi_{bb}\to\Sigma_{b}^{*}}$  & $0.118$ \tabularnewline
		$f_{4}^{\Xi_{bb}\to\Sigma_{b}^{*}}$  & $-0.328$  & $g_{4}^{\Xi_{bb}\to\Sigma_{b}^{*}}$  & $0.087$ \tabularnewline
		\hline 
		$f_{1}^{\Xi_{bb}\to\Xi_{bc}^{*}}$  & $-0.791$  & $g_{1}^{\Xi_{bb}\to\Xi_{bc}^{*}}$  & $-3.044$ \tabularnewline
		$f_{2}^{\Xi_{bb}\to\Xi_{bc}^{*}}$  & $1.284$  & $g_{2}^{\Xi_{bb}\to\Xi_{bc}^{*}}$  & $-0.441$ \tabularnewline
		$f_{3}^{\Xi_{bb}\to\Xi_{bc}^{*}}$  & $-1.287$  & $g_{3}^{\Xi_{bb}\to\Xi_{bc}^{*}}$  & $3.170$ \tabularnewline
		$f_{4}^{\Xi_{bb}\to\Xi_{bc}^{*}}$  & $-1.439$  & $g_{4}^{\Xi_{bb}\to\Xi_{bc}^{*}}$  & $0.342$ \tabularnewline
		\hline 
		$f_{1}^{\Omega_{bb}^{-}\to\Xi_{b}^{\prime*0}}$  & $-0.178$  & $g_{1}^{\Omega_{bb}^{-}\to\Xi_{b}^{\prime*0}}$  & $-0.207$ \tabularnewline
		$f_{2}^{\Omega_{bb}^{-}\to\Xi_{b}^{\prime*0}}$  & $0.225$  & $g_{2}^{\Omega_{bb}^{-}\to\Xi_{b}^{\prime*0}}$  & $0.092$ \tabularnewline
		$f_{3}^{\Omega_{bb}^{-}\to\Xi_{b}^{\prime*0}}$  & $-0.148$  & $g_{3}^{\Omega_{bb}^{-}\to\Xi_{b}^{\prime*0}}$  & $0.104$ \tabularnewline
		$f_{4}^{\Omega_{bb}^{-}\to\Xi_{b}^{\prime*0}}$  & $-0.321$  & $g_{4}^{\Omega_{bb}^{-}\to\Xi_{b}^{\prime*0}}$  & $0.086$ \tabularnewline
		\hline 
		$f_{1}^{\Omega_{bb}^{-}\to\Omega_{bc}^{*0}}$  & $-0.759$  & $g_{1}^{\Omega_{bb}^{-}\to\Omega_{bc}^{*0}}$  & $-2.555$ \tabularnewline
		$f_{2}^{\Omega_{bb}^{-}\to\Omega_{bc}^{*0}}$  & $1.122$  & $g_{2}^{\Omega_{bb}^{-}\to\Omega_{bc}^{*0}}$  & $-0.272$ \tabularnewline
		$f_{3}^{\Omega_{bb}^{-}\to\Omega_{bc}^{*0}}$  & $-1.089$  & $g_{3}^{\Omega_{bb}^{-}\to\Omega_{bc}^{*0}}$  & $2.520$ \tabularnewline
		$f_{4}^{\Omega_{bb}^{-}\to\Omega_{bc}^{*0}}$  & $-1.390$  & $g_{4}^{\Omega_{bb}^{-}\to\Omega_{bc}^{*0}}$  & $0.366$ \tabularnewline
		\hline 
	\end{tabular}
\end{table}

\begin{table}
	\caption{Same as Table \ref{Tab:ff_cc} but for $bc^{(\prime)}$ sector with
		the $c$ quark decay.}
	
	\label{Tab:ff_bc_c}%
	\begin{tabular}{c|r|c|r}
		\hline 
		$F$  & $F(0)$  & $F$  & $F(0)$ \tabularnewline
		\hline 
		$f_{1}^{\Xi_{bc}^{(\prime)+}\to\Sigma_{b}^{*0}}$  & $-1.761$  & $g_{1}^{\Xi_{bc}^{(\prime)+}\to\Sigma_{b}^{*0}}$  & $-16.100$ \tabularnewline
		$f_{2}^{\Xi_{bc}^{(\prime)+}\to\Sigma_{b}^{*0}}$  & $6.031$  & $g_{2}^{\Xi_{bc}^{(\prime)+}\to\Sigma_{b}^{*0}}$  & $4.483$ \tabularnewline
		$f_{3}^{\Xi_{bc}^{(\prime)+}\to\Sigma_{b}^{*0}}$  & $-8.228$  & $g_{3}^{\Xi_{bc}^{(\prime)+}\to\Sigma_{b}^{*0}}$  & $10.280$ \tabularnewline
		$f_{4}^{\Xi_{bc}^{(\prime)+}\to\Sigma_{b}^{*0}}$  & $-3.253$  & $g_{4}^{\Xi_{bc}^{(\prime)+}\to\Sigma_{b}^{*0}}$  & $0.530$ \tabularnewline
		\hline 
		$f_{1}^{\Xi_{bc}^{(\prime)+}\to\Xi_{b}^{\prime*0}}$  & $-2.026$  & $g_{1}^{\Xi_{bc}^{(\prime)+}\to\Xi_{b}^{\prime*0}}$  & $-29.930$ \tabularnewline
		$f_{2}^{\Xi_{bc}^{(\prime)+}\to\Xi_{b}^{\prime*0}}$  & $5.134$  & $g_{2}^{\Xi_{bc}^{(\prime)+}\to\Xi_{b}^{\prime*0}}$  & $3.772$ \tabularnewline
		$f_{3}^{\Xi_{bc}^{(\prime)+}\to\Xi_{b}^{\prime*0}}$  & $-9.413$  & $g_{3}^{\Xi_{bc}^{(\prime)+}\to\Xi_{b}^{\prime*0}}$  & $24.170$ \tabularnewline
		$f_{4}^{\Xi_{bc}^{(\prime)+}\to\Xi_{b}^{\prime*0}}$  & $-3.741$  & $g_{4}^{\Xi_{bc}^{(\prime)+}\to\Xi_{b}^{\prime*0}}$  & $0.546$ \tabularnewline
		\hline 
		$f_{1}^{\Xi_{bc}^{(\prime)0}\to\Sigma_{b}^{*-}}$  & $-1.768$  & $g_{1}^{\Xi_{bc}^{(\prime)0}\to\Sigma_{b}^{*-}}$  & $-16.360$ \tabularnewline
		$f_{2}^{\Xi_{bc}^{(\prime)0}\to\Sigma_{b}^{*-}}$  & $6.140$  & $g_{2}^{\Xi_{bc}^{(\prime)0}\to\Sigma_{b}^{*-}}$  & $4.451$ \tabularnewline
		$f_{3}^{\Xi_{bc}^{(\prime)0}\to\Sigma_{b}^{*-}}$  & $-8.360$  & $g_{3}^{\Xi_{bc}^{(\prime)0}\to\Sigma_{b}^{*-}}$  & $10.570$ \tabularnewline
		$f_{4}^{\Xi_{bc}^{(\prime)0}\to\Sigma_{b}^{*-}}$  & $-3.264$  & $g_{4}^{\Xi_{bc}^{(\prime)0}\to\Sigma_{b}^{*-}}$  & $0.527$ \tabularnewline
		\hline 
		$f_{1}^{\Xi_{bc}^{(\prime)0}\to\Xi_{b}^{\prime*-}}$  & $-2.052$  & $g_{1}^{\Xi_{bc}^{(\prime)0}\to\Xi_{b}^{\prime*-}}$  & $-31.380$ \tabularnewline
		$f_{2}^{\Xi_{bc}^{(\prime)0}\to\Xi_{b}^{\prime*-}}$  & $5.597$  & $g_{2}^{\Xi_{bc}^{(\prime)0}\to\Xi_{b}^{\prime*-}}$  & $3.552$ \tabularnewline
		$f_{3}^{\Xi_{bc}^{(\prime)0}\to\Xi_{b}^{\prime*-}}$  & $-10.010$  & $g_{3}^{\Xi_{bc}^{(\prime)0}\to\Xi_{b}^{\prime*-}}$  & $25.820$ \tabularnewline
		$f_{4}^{\Xi_{bc}^{(\prime)0}\to\Xi_{b}^{\prime*-}}$  & $-3.785$  & $g_{4}^{\Xi_{bc}^{(\prime)0}\to\Xi_{b}^{\prime*-}}$  & $0.535$ \tabularnewline
		\hline 
		$f_{1}^{\Omega_{bc}^{(\prime)0}\to\Xi_{b}^{\prime*-}}$  & $-2.043$  & $g_{1}^{\Omega_{bc}^{(\prime)0}\to\Xi_{b}^{\prime*-}}$  & $-27.570$ \tabularnewline
		$f_{2}^{\Omega_{bc}^{(\prime)0}\to\Xi_{b}^{\prime*-}}$  & $11.110$  & $g_{2}^{\Omega_{bc}^{(\prime)0}\to\Xi_{b}^{\prime*-}}$  & $3.368$ \tabularnewline
		$f_{3}^{\Omega_{bc}^{(\prime)0}\to\Xi_{b}^{\prime*-}}$  & $-14.160$  & $g_{3}^{\Omega_{bc}^{(\prime)0}\to\Xi_{b}^{\prime*-}}$  & $22.720$ \tabularnewline
		$f_{4}^{\Omega_{bc}^{(\prime)0}\to\Xi_{b}^{\prime*-}}$  & $-3.740$  & $g_{4}^{\Omega_{bc}^{(\prime)0}\to\Xi_{b}^{\prime*-}}$  & $0.440$ \tabularnewline
		\hline 
		$f_{1}^{\Omega_{bc}^{(\prime)0}\to\Omega_{b}^{*-}}$  & $-2.465$  & $g_{1}^{\Omega_{bc}^{(\prime)0}\to\Omega_{b}^{*-}}$  & $-57.090$ \tabularnewline
		$f_{2}^{\Omega_{bc}^{(\prime)0}\to\Omega_{b}^{*-}}$  & $14.850$  & $g_{2}^{\Omega_{bc}^{(\prime)0}\to\Omega_{b}^{*-}}$  & $0.132$ \tabularnewline
		$f_{3}^{\Omega_{bc}^{(\prime)0}\to\Omega_{b}^{*-}}$  & $-21.280$  & $g_{3}^{\Omega_{bc}^{(\prime)0}\to\Omega_{b}^{*-}}$  & $54.650$ \tabularnewline
		$f_{4}^{\Omega_{bc}^{(\prime)0}\to\Omega_{b}^{*-}}$  & $-4.494$  & $g_{4}^{\Omega_{bc}^{(\prime)0}\to\Omega_{b}^{*-}}$  & $0.386$ \tabularnewline
		\hline 
	\end{tabular}
\end{table}

\begin{table}
	\caption{Same as Table \ref{Tab:ff_bb} but for the $bc^{(\prime)}$ sector
		with the $b$ quark decay. }
	\label{Tab:ff_bc_b}%
	\begin{tabular}{c|r|c|r}
		\hline 
		$F$  & $F(0)$  & $F$  & $F(0)$ \tabularnewline
		\hline 
		$f_{1}^{\Xi_{bc}^{(\prime)}\to\Sigma_{c}^{*}}$  & $-0.114$  & $g_{1}^{\Xi_{bc}^{(\prime)}\to\Sigma_{c}^{*}}$  & $-0.024$ \tabularnewline
		$f_{2}^{\Xi_{bc}^{(\prime)}\to\Sigma_{c}^{*}}$  & $0.040$  & $g_{2}^{\Xi_{bc}^{(\prime)}\to\Sigma_{c}^{*}}$  & $0.062$ \tabularnewline
		$f_{3}^{\Xi_{bc}^{(\prime)}\to\Sigma_{c}^{*}}$  & $0.030$  & $g_{3}^{\Xi_{bc}^{(\prime)}\to\Sigma_{c}^{*}}$  & $-0.069$ \tabularnewline
		$f_{4}^{\Xi_{bc}^{(\prime)}\to\Sigma_{c}^{*}}$  & $-0.239$  & $g_{4}^{\Xi_{bc}^{(\prime)}\to\Sigma_{c}^{*}}$  & $0.156$ \tabularnewline
		\hline 
		$f_{1}^{\Xi_{bc}^{(\prime)}\to\Xi_{cc}^{*}}$  & $-0.497$  & $g_{1}^{\Xi_{bc}^{(\prime)}\to\Xi_{cc}^{*}}$  & $-0.446$ \tabularnewline
		$f_{2}^{\Xi_{bc}^{(\prime)}\to\Xi_{cc}^{*}}$  & $0.265$  & $g_{2}^{\Xi_{bc}^{(\prime)}\to\Xi_{cc}^{*}}$  & $0.118$ \tabularnewline
		$f_{3}^{\Xi_{bc}^{(\prime)}\to\Xi_{cc}^{*}}$  & $-0.011$  & $g_{3}^{\Xi_{bc}^{(\prime)}\to\Xi_{cc}^{*}}$  & $0.117$ \tabularnewline
		$f_{4}^{\Xi_{bc}^{(\prime)}\to\Xi_{cc}^{*}}$  & $-0.969$  & $g_{4}^{\Xi_{bc}^{(\prime)}\to\Xi_{cc}^{*}}$  & $0.539$ \tabularnewline
		\hline 
		$f_{1}^{\Omega_{bc}^{(\prime)0}\to\Xi_{c}^{\prime*+}}$  & $-0.112$  & $g_{1}^{\Omega_{bc}^{(\prime)0}\to\Xi_{c}^{\prime*+}}$  & $-0.028$ \tabularnewline
		$f_{2}^{\Omega_{bc}^{(\prime)0}\to\Xi_{c}^{\prime*+}}$  & $0.043$  & $g_{2}^{\Omega_{bc}^{(\prime)0}\to\Xi_{c}^{\prime*+}}$  & $0.061$ \tabularnewline
		$f_{3}^{\Omega_{bc}^{(\prime)0}\to\Xi_{c}^{\prime*+}}$  & $0.025$  & $g_{3}^{\Omega_{bc}^{(\prime)0}\to\Xi_{c}^{\prime*+}}$  & $-0.059$ \tabularnewline
		$f_{4}^{\Omega_{bc}^{(\prime)0}\to\Xi_{c}^{\prime*+}}$  & $-0.231$  & $g_{4}^{\Omega_{bc}^{(\prime)0}\to\Xi_{c}^{\prime*+}}$  & $0.140$ \tabularnewline
		\hline 
		$f_{1}^{\Omega_{bc}^{(\prime)0}\to\Omega_{cc}^{*+}}$  & $-0.519$  & $g_{1}^{\Omega_{bc}^{(\prime)0}\to\Omega_{cc}^{*+}}$  & $-0.614$ \tabularnewline
		$f_{2}^{\Omega_{bc}^{(\prime)0}\to\Omega_{cc}^{*+}}$  & $0.314$  & $g_{2}^{\Omega_{bc}^{(\prime)0}\to\Omega_{cc}^{*+}}$  & $0.061$ \tabularnewline
		$f_{3}^{\Omega_{bc}^{(\prime)0}\to\Omega_{cc}^{*+}}$  & $-0.059$  & $g_{3}^{\Omega_{bc}^{(\prime)0}\to\Omega_{cc}^{*+}}$  & $0.356$ \tabularnewline
		$f_{4}^{\Omega_{bc}^{(\prime)0}\to\Omega_{cc}^{*+}}$  & $-1.003$  & $g_{4}^{\Omega_{bc}^{(\prime)0}\to\Omega_{cc}^{*+}}$  & $0.497$ \tabularnewline
		\hline 
	\end{tabular}
\end{table}

\subsection{Results for semi-leptonic decays}

Helicity amplitudes are defined by $H_{\lambda^{\prime},\lambda_{W}}^{V}\equiv\langle{\cal B}_{f}^{*}(\lambda^{\prime})|\bar{q}\gamma^{\mu}Q|{\cal B}_{i}(\lambda)\rangle\epsilon_{W\mu}^{*}(\lambda_{W})$
and $H_{\lambda^{\prime},\lambda_{W}}^{A}\equiv\langle{\cal B}_{f}^{*}(\lambda^{\prime})|\bar{q}\gamma^{\mu}\gamma_{5}Q|{\cal B}_{i}(\lambda)\rangle\epsilon_{W\mu}^{*}(\lambda_{W})$
respectively, where $\lambda=\lambda_{W}-\lambda^{\prime}$ is understood.
These helicity amplitudes are related to the form factors by the following
expressions.
\begin{eqnarray}
H_{3/2,1}^{V,A} & = & \mp i\sqrt{2MM^{\prime}(\omega\mp1)}f_{4}^{V,A},\\
H_{1/2,1}^{V,A} & = & i\sqrt{\frac{2}{3}}\sqrt{MM^{\prime}(\omega\mp1)}\left[f_{4}^{V,A}-2(\omega\pm1)f_{1}^{V,A}\right],\\
H_{1/2,0}^{V,A} & = & \pm i\frac{1}{\sqrt{q^{2}}}\frac{2}{\sqrt{3}}\sqrt{MM^{\prime}(\omega\mp1)}\Big[(M\omega-M^{\prime})f_{4}^{V,A}\mp(M\mp M^{\prime})(\omega\pm1)f_{1}^{V,A}\nonumber \\
&  & \qquad\qquad\qquad\qquad\qquad\qquad+M^{\prime}(\omega^{2}-1)f_{2}^{V,A}+M(\omega^{2}-1)f_{3}^{V,A}\Big],
\end{eqnarray}
where the upper (lower) sign corresponds to $V$ ($A$), $f_{i}^{V}=f_{i}$
($f_{i}^{A}=g_{i}$), $\omega$ is defined in Eq. (\ref{eq:omega}), $M$ ($M^{\prime}$) is the mass of the baryon in the initial (final) state. The remaining helicity amplitudes can be obtained
by
\begin{equation}
H_{-\lambda^{\prime},-\lambda_{W}}^{V,A}=\mp H_{\lambda^{\prime},\lambda_{W}}^{V,A}.
\end{equation}
Partial differential decay widths are obtained as 
\begin{eqnarray}
\frac{d\Gamma_{T}}{d\omega} & = & \frac{G_{F}^{2}}{(2\pi)^{3}}|V_{{\rm CKM}}|^{2}\frac{q^{2}M^{\prime2}\sqrt{\omega^{2}-1}}{12M}[|H_{1/2,1}|^{2}+|H_{-1/2,-1}|^{2}+|H_{3/2,1}|^{2}+|H_{-3/2,-1}|^{2}],\\
\frac{d\Gamma_{L}}{d\omega} & = & \frac{G_{F}^{2}}{(2\pi)^{3}}|V_{{\rm CKM}}|^{2}\frac{q^{2}M^{\prime2}\sqrt{\omega^{2}-1}}{12M}[|H_{1/2,0}|^{2}+|H_{-1/2,0}|^{2}].
\end{eqnarray}

Numerical results are collected in Tables \ref{Tab:semi_cc} to
\ref{Tab:semi_bcp_b}. Some comments are given in subsection \ref{subsec:some_other_discussions}.

\begin{table}
	\caption{Semi-leptonic decays for the $cc$ sector.}
	\label{Tab:semi_cc}%
	\begin{tabular}{l|c|c|c}
		\hline 
		channels  & $\Gamma/\text{~GeV}$  & ${\cal B}$  & $\Gamma_{L}/\Gamma_{T}$ \tabularnewline
		\hline 
		$\Xi_{cc}^{++}\to\Sigma_{c}^{*+}e^{+}\nu_{e}$  & $1.26\times10^{-15}$  & $5.73\times10^{-4}$  & $0.85$\tabularnewline
		$\Xi_{cc}^{+}\to\Sigma_{c}^{*0}e^{+}\nu_{e}$  & $2.51\times10^{-15}$  & $3.82\times10^{-4}$  & $0.85$\tabularnewline
		$\Omega_{cc}^{+}\to\Xi_{c}^{\prime*0}e^{+}\nu_{e}$  & $1.19\times10^{-15}$  & $1.82\times10^{-4}$  & $0.87$\tabularnewline
		$\Xi_{cc}^{++}\to\Xi_{c}^{\prime*+}e^{+}\nu_{e}$  & $1.61\times10^{-14}$  & $7.34\times10^{-3}$  & $0.99$\tabularnewline
		$\Xi_{cc}^{+}\to\Xi_{c}^{\prime*0}e^{+}\nu_{e}$  & $1.61\times10^{-14}$  & $2.45\times10^{-3}$  & $0.99$\tabularnewline
		$\Omega_{cc}^{+}\to\Omega_{c}^{*0}e^{+}\nu_{e}$  & $3.20\times10^{-14}$  & $4.87\times10^{-3}$  & $0.99$\tabularnewline
		\hline 
	\end{tabular}
\end{table}

\begin{table}
	\caption{Semi-leptonic decays for the $bb$ sector.}
	\label{Tab:semi_bb}%
	\begin{tabular}{l|c|c|c}
		\hline 
		channels  & $\Gamma/\text{~GeV}$  & ${\cal B}$  & $\Gamma_{L}/\Gamma_{T}$ \tabularnewline
		\hline 
		$\Xi_{bb}^{0}\to\Sigma_{b}^{*+}e^{-}\bar{\nu}_{e}$  & $3.88\times10^{-17}$  & $2.18\times10^{-5}$  & $0.85$\tabularnewline
		$\Xi_{bb}^{-}\to\Sigma_{b}^{*0}e^{-}\bar{\nu}_{e}$  & $1.94\times10^{-17}$  & $1.09\times10^{-5}$  & $0.85$\tabularnewline
		$\Omega_{bb}^{-}\to\Xi_{b}^{\prime*0}e^{-}\bar{\nu}_{e}$  & $1.90\times10^{-17}$  & $2.32\times10^{-5}$  & $0.84$\tabularnewline
		$\Xi_{bb}^{0}\to\Xi_{bc}^{*+}e^{-}\bar{\nu}_{e}$  & $6.37\times10^{-15}$  & $3.58\times10^{-3}$  & $1.43$\tabularnewline
		$\Xi_{bb}^{-}\to\Xi_{bc}^{*0}e^{-}\bar{\nu}_{e}$  & $6.37\times10^{-15}$  & $3.58\times10^{-3}$  & $1.43$\tabularnewline
		$\Omega_{bb}^{-}\to\Omega_{bc}^{*0}e^{-}\bar{\nu}_{e}$  & $7.03\times10^{-15}$  & $8.55\times10^{-3}$  & $1.31$\tabularnewline
		\hline 
	\end{tabular}
\end{table}

\begin{table}
	\caption{Semi-leptonic decays for the $bc$ sector with the $c$ quark decay.}
	\label{Tab:semi_bc_c}%
	\begin{tabular}{l|c|c|c}
		\hline 
		channels  & $\Gamma/\text{~GeV}$  & ${\cal B}$  & $\Gamma_{L}/\Gamma_{T}$ \tabularnewline
		\hline 
		$\Xi_{bc}^{+}\to\Sigma_{b}^{*0}e^{+}\nu_{e}$  & $1.10\times10^{-15}$  & $4.07\times10^{-4}$  & $0.69$\tabularnewline
		$\Xi_{bc}^{0}\to\Sigma_{b}^{*-}e^{+}\nu_{e}$  & $2.17\times10^{-15}$  & $3.06\times10^{-4}$  & $0.69$\tabularnewline
		$\Omega_{bc}^{0}\to\Xi_{b}^{\prime*-}e^{+}\nu_{e}$  & $6.98\times10^{-16}$  & $2.33\times10^{-4}$  & $0.80$\tabularnewline
		$\Xi_{bc}^{+}\to\Xi_{b}^{\prime*0}e^{+}\nu_{e}$  & $1.33\times10^{-14}$  & $4.95\times10^{-3}$  & $0.77$\tabularnewline
		$\Xi_{bc}^{0}\to\Xi_{b}^{\prime*-}e^{+}\nu_{e}$  & $1.27\times10^{-14}$  & $1.80\times10^{-3}$  & $0.78$\tabularnewline
		$\Omega_{bc}^{0}\to\Omega_{b}^{*-}e^{+}\nu_{e}$  & $1.47\times10^{-14}$  & $4.91\times10^{-3}$  & $0.97$\tabularnewline
		\hline 
	\end{tabular}
\end{table}

\begin{table}
	\caption{Semi-leptonic decays for the $bc$ sector with the $b$ quark decay.}
	\label{Tab:semi_bc_b}%
	\begin{tabular}{l|c|c|c}
		\hline 
		channels  & $\Gamma/\text{~GeV}$  & ${\cal B}$  & $\Gamma_{L}/\Gamma_{T}$ \tabularnewline
		\hline 
		$\Xi_{bc}^{+}\to\Sigma_{c}^{*++}e^{-}\bar{\nu}_{e}$  & $3.32\times10^{-17}$  & $1.23\times10^{-5}$  & $0.81$\tabularnewline
		$\Xi_{bc}^{0}\to\Sigma_{c}^{*+}e^{-}\bar{\nu}_{e}$  & $1.66\times10^{-17}$  & $2.35\times10^{-6}$  & $0.81$\tabularnewline
		$\Omega_{bc}^{0}\to\Xi_{c}^{\prime*+}e^{-}\bar{\nu}_{e}$  & $1.26\times10^{-17}$  & $4.23\times10^{-6}$  & $0.84$\tabularnewline
		$\Xi_{bc}^{+}\to\Xi_{cc}^{*++}e^{-}\bar{\nu}_{e}$  & $8.96\times10^{-15}$  & $3.32\times10^{-3}$  & $1.18$\tabularnewline
		$\Xi_{bc}^{0}\to\Xi_{cc}^{*+}e^{-}\bar{\nu}_{e}$  & $8.96\times10^{-15}$  & $1.27\times10^{-3}$  & $1.18$\tabularnewline
		$\Omega_{bc}^{0}\to\Omega_{cc}^{*+}e^{-}\bar{\nu}_{e}$  & $7.53\times10^{-15}$  & $2.52\times10^{-3}$  & $1.28$\tabularnewline
		\hline 
	\end{tabular}
\end{table}

\begin{table}
	\caption{Semi-leptonic decays for the $bc^{\prime}$ sector with the $c$ quark
		decay.}
	\label{Tab:semi_bcp_c}%
	\begin{tabular}{l|c|c|c}
		\hline 
		channels  & $\Gamma/\text{~GeV}$  & ${\cal B}$  & $\Gamma_{L}/\Gamma_{T}$ \tabularnewline
		\hline 
		$\Xi_{bc}^{\prime+}\to\Sigma_{b}^{*0}e^{+}\nu_{e}$  & $3.30\times10^{-15}$  & $1.22\times10^{-3}$  & $0.69$\tabularnewline
		$\Xi_{bc}^{\prime0}\to\Sigma_{b}^{*-}e^{+}\nu_{e}$  & $6.50\times10^{-15}$  & $9.18\times10^{-4}$  & $0.69$\tabularnewline
		$\Omega_{bc}^{\prime0}\to\Xi_{b}^{\prime*-}e^{+}\nu_{e}$  & $2.09\times10^{-15}$  & $7.00\times10^{-4}$  & $0.80$\tabularnewline
		$\Xi_{bc}^{\prime+}\to\Xi_{b}^{\prime*0}e^{+}\nu_{e}$  & $4.00\times10^{-14}$  & $1.48\times10^{-2}$  & $0.77$\tabularnewline
		$\Xi_{bc}^{\prime0}\to\Xi_{b}^{\prime*-}e^{+}\nu_{e}$  & $3.82\times10^{-14}$  & $5.40\times10^{-3}$  & $0.78$\tabularnewline
		$\Omega_{bc}^{\prime0}\to\Omega_{b}^{*-}e^{+}\nu_{e}$  & $4.40\times10^{-14}$  & $1.47\times10^{-2}$  & $0.97$\tabularnewline
		\hline 
	\end{tabular}
\end{table}

\begin{table}
	\caption{Semi-leptonic decays for the $bc^{\prime}$ sector with the $b$ quark
		decay.}
	\label{Tab:semi_bcp_b}%
	\begin{tabular}{l|c|c|c}
		\hline 
		channels  & $\Gamma/\text{~GeV}$  & ${\cal B}$  & $\Gamma_{L}/\Gamma_{T}$ \tabularnewline
		\hline 
		$\Xi_{bc}^{\prime+}\to\Sigma_{c}^{*++}e^{-}\bar{\nu}_{e}$  & $9.97\times10^{-17}$  & $3.70\times10^{-5}$  & $0.81$\tabularnewline
		$\Xi_{bc}^{\prime0}\to\Sigma_{c}^{*+}e^{-}\bar{\nu}_{e}$  & $4.99\times10^{-17}$  & $7.05\times10^{-6}$  & $0.81$\tabularnewline
		$\Omega_{bc}^{\prime0}\to\Xi_{c}^{\prime*+}e^{-}\bar{\nu}_{e}$  & $3.79\times10^{-17}$  & $1.27\times10^{-5}$  & $0.84$\tabularnewline
		$\Xi_{bc}^{\prime+}\to\Xi_{cc}^{*++}e^{-}\bar{\nu}_{e}$  & $2.69\times10^{-14}$  & $9.97\times10^{-3}$  & $1.18$\tabularnewline
		$\Xi_{bc}^{\prime0}\to\Xi_{cc}^{*+}e^{-}\bar{\nu}_{e}$  & $2.69\times10^{-14}$  & $3.80\times10^{-3}$  & $1.18$\tabularnewline
		$\Omega_{bc}^{\prime0}\to\Omega_{cc}^{*+}e^{-}\bar{\nu}_{e}$  & $2.26\times10^{-14}$  & $7.55\times10^{-3}$  & $1.28$\tabularnewline
		\hline 
	\end{tabular}
\end{table}

\subsection{Results for nonleptonic decays}

For the nonleptonic processes, we are constrained to consider only those of a W boson emitting outward. For the process with a pseudoscalar meson in the final
state, the decay width is obtained as
\begin{equation}
	\Gamma=|\lambda|^{2}f_{P}^{2}\frac{M|\vec{P}^{\prime}|^{3}}{6\pi M^{\prime}}[(\omega-1)(B^{2}-2AB)+2A^{2}\omega],
\end{equation}
with 
\begin{eqnarray}
	\lambda & \equiv & \frac{G_{F}}{\sqrt{2}}V_{cb}V_{q_{1}q_{2}}^{*}a_{1},\\
	A & = & (M-M^{\prime})\frac{g_{1}}{M}+\frac{g_{2}}{M^{2}}(P\cdot q)+\frac{g_{3}}{MM^{\prime}}(P^{\prime}\cdot q)+g_{4},\\
	B & = & -(M+M^{\prime})\frac{f_{1}}{M}+\frac{f_{2}}{M^{2}}(P\cdot q)+\frac{f_{3}}{MM^{\prime}}(P^{\prime}\cdot q)+f_{4}.
\end{eqnarray}
Here $a_{1}\equiv C_{1}+C_{2}/3$.

For the process with a vector meson in the final state,
the decay width is obtained as
\begin{eqnarray}
	\Gamma & = & |\lambda|^{2}f_{V}^{2}m^{2}\frac{|\vec{P}^{\prime}|}{16\pi M^{2}}[|H_{1/2,1}|^{2}+|H_{-1/2,-1}|^{2}+|H_{3/2,1}|^{2}+|H_{-3/2,-1}|^{2}\nonumber \\
	&  & \quad\quad\quad\quad\quad\quad\quad+|H_{1/2,0}|^{2}+|H_{-1/2,0}|^{2}].
\end{eqnarray}
Note that in the above equations, $q^{2}=m^{2}$ is understood, where
$m$ is the mass of the meson.

All the corresponding results are collected in Tables \ref{Tab:non_cc}
to \ref{Tab:non_bcp_b}. Some comments are given in subsection \ref{subsec:some_other_discussions}.

\begin{table}
	\caption{Nonleptonic decays for $cc$ sector.}
	\label{Tab:non_cc}%
	\begin{tabular}{l|c|c|l|c|c}
		\hline 
		channels  & $\Gamma/\text{~GeV}$  & ${\cal B}$  & channels  & $\Gamma/\text{~GeV}$ & ${\cal B}$ \tabularnewline
		\hline 
		$\Xi_{cc}^{++}\to\Sigma_{c}^{*+}\pi^{+}$  & $1.16\times10^{-15}$  & $5.28\times10^{-4}$  & $\Xi_{cc}^{++}\to\Sigma_{c}^{*+}\rho^{+}$  & $3.78\times10^{-15}$  & $1.73\times10^{-3}$ \tabularnewline
		$\Xi_{cc}^{++}\to\Sigma_{c}^{*+}K^{*+}$  & $1.60\times10^{-16}$  & $7.31\times10^{-5}$  & $\Xi_{cc}^{++}\to\Sigma_{c}^{*+}K^{+}$  & $4.97\times10^{-17}$  & $2.27\times10^{-5}$ \tabularnewline
		\hline 
		$\Xi_{cc}^{++}\to\Xi_{c}^{\prime*+}\pi^{+}$  & $2.24\times10^{-14}$  & $1.02\times10^{-2}$  & $\Xi_{cc}^{++}\to\Xi_{c}^{\prime*+}\rho^{+}$  & $4.85\times10^{-14}$  & $2.21\times10^{-2}$ \tabularnewline
		$\Xi_{cc}^{++}\to\Xi_{c}^{\prime*+}K^{*+}$  & $1.32\times10^{-15}$  & $6.03\times10^{-4}$  & $\Xi_{cc}^{++}\to\Xi_{c}^{\prime*+}K^{+}$  & $7.05\times10^{-16}$  & $3.21\times10^{-4}$ \tabularnewline
		\hline 
		$\Xi_{cc}^{+}\to\Sigma_{c}^{*0}\pi^{+}$  & $2.32\times10^{-15}$  & $3.52\times10^{-4}$  & $\Xi_{cc}^{+}\to\Sigma_{c}^{*0}\rho^{+}$  & $7.57\times10^{-15}$  & $1.15\times10^{-3}$ \tabularnewline
		$\Xi_{cc}^{+}\to\Sigma_{c}^{*0}K^{*+}$  & $3.21\times10^{-16}$  & $4.87\times10^{-5}$  & $\Xi_{cc}^{+}\to\Sigma_{c}^{*0}K^{+}$  & $9.95\times10^{-17}$  & $1.51\times10^{-5}$ \tabularnewline
		\hline 
		$\Xi_{cc}^{+}\to\Xi_{c}^{\prime*0}\pi^{+}$  & $2.24\times10^{-14}$  & $3.40\times10^{-3}$  & $\Xi_{cc}^{+}\to\Xi_{c}^{\prime*0}\rho^{+}$  & $4.85\times10^{-14}$  & $7.37\times10^{-3}$ \tabularnewline
		$\Xi_{cc}^{+}\to\Xi_{c}^{\prime*0}K^{*+}$  & $1.32\times10^{-15}$  & $2.01\times10^{-4}$  & $\Xi_{cc}^{+}\to\Xi_{c}^{\prime*0}K^{+}$  & $7.05\times10^{-16}$  & $1.07\times10^{-4}$ \tabularnewline
		\hline 
		$\Omega_{cc}^{+}\to\Xi_{c}^{\prime*0}\pi^{+}$  & $1.10\times10^{-15}$  & $1.68\times10^{-4}$  & $\Omega_{cc}^{+}\to\Xi_{c}^{\prime*0}\rho^{+}$  & $3.64\times10^{-15}$  & $5.53\times10^{-4}$ \tabularnewline
		$\Omega_{cc}^{+}\to\Xi_{c}^{\prime*0}K^{*+}$  & $1.53\times10^{-16}$  & $2.32\times10^{-5}$  & $\Omega_{cc}^{+}\to\Xi_{c}^{\prime*0}K^{+}$  & $4.74\times10^{-17}$  & $7.20\times10^{-6}$ \tabularnewline
		\hline 
		$\Omega_{cc}^{+}\to\Omega_{c}^{*0}\pi^{+}$  & $4.26\times10^{-14}$  & $6.47\times10^{-3}$  & $\Omega_{cc}^{+}\to\Omega_{c}^{*0}\rho^{+}$  & $9.88\times10^{-14}$  & $1.50\times10^{-2}$ \tabularnewline
		$\Omega_{cc}^{+}\to\Omega_{c}^{*0}K^{*+}$  & $2.79\times10^{-15}$  & $4.24\times10^{-4}$  & $\Omega_{cc}^{+}\to\Omega_{c}^{*0}K^{+}$  & $1.38\times10^{-15}$  & $2.10\times10^{-4}$ \tabularnewline
		\hline 
	\end{tabular}
\end{table}

\begin{table}
	\caption{Nonleptonic decays for $bb$ sector.}
	\label{Tab:non_bb}%
	\begin{tabular}{l|c|c|l|c|c}
		\hline 
		channels  & $\Gamma/\text{~GeV}$  & ${\cal B}$  & channels  & $\Gamma/\text{~GeV}$ & ${\cal B}$ \tabularnewline
		\hline 
		$\Xi_{bb}^{0}\to\Sigma_{b}^{*+}\pi^{-}$  & $1.26\times10^{-18}$  & $7.06\times10^{-7}$  & $\Xi_{bb}^{0}\to\Sigma_{b}^{*+}\rho^{-}$  & $3.24\times10^{-18}$  & $1.82\times10^{-6}$ \tabularnewline
		$\Xi_{bb}^{0}\to\Sigma_{b}^{*+}a_{1}^{-}$  & $4.38\times10^{-18}$  & $2.47\times10^{-6}$  & $\Xi_{bb}^{0}\to\Sigma_{b}^{*+}K^{-}$  & $9.99\times10^{-20}$  & $5.62\times10^{-8}$ \tabularnewline
		$\Xi_{bb}^{0}\to\Sigma_{b}^{*+}K^{*-}$  & $1.67\times10^{-19}$  & $9.41\times10^{-8}$  & $\Xi_{bb}^{0}\to\Sigma_{b}^{*+}D^{-}$  & $1.44\times10^{-19}$  & $8.12\times10^{-8}$ \tabularnewline
		$\Xi_{bb}^{0}\to\Sigma_{b}^{*+}D^{*-}$  & $2.58\times10^{-19}$  & $1.45\times10^{-7}$  & $\Xi_{bb}^{0}\to\Sigma_{b}^{*+}D_{s}^{-}$  & $3.77\times10^{-18}$  & $2.12\times10^{-6}$ \tabularnewline
		$\Xi_{bb}^{0}\to\Sigma_{b}^{*+}D_{s}^{*-}$  & $6.34\times10^{-18}$  & $3.57\times10^{-6}$  &  &  & \tabularnewline
		\hline 
		$\Xi_{bb}^{0}\to\Xi_{bc}^{*+}\pi^{-}$  & $8.21\times10^{-16}$  & $4.62\times10^{-4}$  & $\Xi_{bb}^{0}\to\Xi_{bc}^{*+}\rho^{-}$  & $2.19\times10^{-15}$  & $1.23\times10^{-3}$ \tabularnewline
		$\Xi_{bb}^{0}\to\Xi_{bc}^{*+}a_{1}^{-}$  & $2.76\times10^{-15}$  & $1.55\times10^{-3}$  & $\Xi_{bb}^{0}\to\Xi_{bc}^{*+}K^{-}$  & $6.28\times10^{-17}$  & $3.53\times10^{-5}$ \tabularnewline
		$\Xi_{bb}^{0}\to\Xi_{bc}^{*+}K^{*-}$  & $1.12\times10^{-16}$  & $6.28\times10^{-5}$  & $\Xi_{bb}^{0}\to\Xi_{bc}^{*+}D^{-}$  & $4.50\times10^{-17}$  & $2.53\times10^{-5}$ \tabularnewline
		$\Xi_{bb}^{0}\to\Xi_{bc}^{*+}D^{*-}$  & $1.26\times10^{-16}$  & $7.11\times10^{-5}$  & $\Xi_{bb}^{0}\to\Xi_{bc}^{*+}D_{s}^{-}$  & $1.06\times10^{-15}$  & $5.97\times10^{-4}$ \tabularnewline
		$\Xi_{bb}^{0}\to\Xi_{bc}^{*+}D_{s}^{*-}$  & $2.95\times10^{-15}$  & $1.66\times10^{-3}$  &  &  & \tabularnewline
		\hline 
		$\Xi_{bb}^{-}\to\Sigma_{b}^{*0}\pi^{-}$  & $6.28\times10^{-19}$  & $3.53\times10^{-7}$  & $\Xi_{bb}^{-}\to\Sigma_{b}^{*0}\rho^{-}$  & $1.62\times10^{-18}$  & $9.11\times10^{-7}$ \tabularnewline
		$\Xi_{bb}^{-}\to\Sigma_{b}^{*0}a_{1}^{-}$  & $2.19\times10^{-18}$  & $1.23\times10^{-6}$  & $\Xi_{bb}^{-}\to\Sigma_{b}^{*0}K^{-}$  & $5.00\times10^{-20}$  & $2.81\times10^{-8}$ \tabularnewline
		$\Xi_{bb}^{-}\to\Sigma_{b}^{*0}K^{*-}$  & $8.37\times10^{-20}$  & $4.71\times10^{-8}$  & $\Xi_{bb}^{-}\to\Sigma_{b}^{*0}D^{-}$  & $7.21\times10^{-20}$  & $4.06\times10^{-8}$ \tabularnewline
		$\Xi_{bb}^{-}\to\Sigma_{b}^{*0}D^{*-}$  & $1.29\times10^{-19}$  & $7.26\times10^{-8}$  & $\Xi_{bb}^{-}\to\Sigma_{b}^{*0}D_{s}^{-}$  & $1.88\times10^{-18}$  & $1.06\times10^{-6}$ \tabularnewline
		$\Xi_{bb}^{-}\to\Sigma_{b}^{*0}D_{s}^{*-}$  & $3.17\times10^{-18}$  & $1.78\times10^{-6}$  &  &  & \tabularnewline
		\hline 
		$\Xi_{bb}^{-}\to\Xi_{bc}^{*0}\pi^{-}$  & $8.21\times10^{-16}$  & $4.62\times10^{-4}$  & $\Xi_{bb}^{-}\to\Xi_{bc}^{*0}\rho^{-}$  & $2.19\times10^{-15}$  & $1.23\times10^{-3}$ \tabularnewline
		$\Xi_{bb}^{-}\to\Xi_{bc}^{*0}a_{1}^{-}$  & $2.76\times10^{-15}$  & $1.55\times10^{-3}$  & $\Xi_{bb}^{-}\to\Xi_{bc}^{*0}K^{-}$  & $6.28\times10^{-17}$  & $3.53\times10^{-5}$ \tabularnewline
		$\Xi_{bb}^{-}\to\Xi_{bc}^{*0}K^{*-}$  & $1.12\times10^{-16}$  & $6.28\times10^{-5}$  & $\Xi_{bb}^{-}\to\Xi_{bc}^{*0}D^{-}$  & $4.50\times10^{-17}$  & $2.53\times10^{-5}$ \tabularnewline
		$\Xi_{bb}^{-}\to\Xi_{bc}^{*0}D^{*-}$  & $1.26\times10^{-16}$  & $7.11\times10^{-5}$  & $\Xi_{bb}^{-}\to\Xi_{bc}^{*0}D_{s}^{-}$  & $1.06\times10^{-15}$  & $5.97\times10^{-4}$ \tabularnewline
		$\Xi_{bb}^{-}\to\Xi_{bc}^{*0}D_{s}^{*-}$  & $2.95\times10^{-15}$  & $1.66\times10^{-3}$  &  &  & \tabularnewline
		\hline 
		$\Omega_{bb}^{-}\to\Xi_{b}^{\prime*0}\pi^{-}$  & $5.85\times10^{-19}$  & $7.11\times10^{-7}$  & $\Omega_{bb}^{-}\to\Xi_{b}^{\prime*0}\rho^{-}$  & $1.52\times10^{-18}$  & $1.85\times10^{-6}$ \tabularnewline
		$\Omega_{bb}^{-}\to\Xi_{b}^{\prime*0}a_{1}^{-}$  & $2.06\times10^{-18}$  & $2.51\times10^{-6}$  & $\Omega_{bb}^{-}\to\Xi_{b}^{\prime*0}K^{-}$  & $4.66\times10^{-20}$  & $5.67\times10^{-8}$ \tabularnewline
		$\Omega_{bb}^{-}\to\Xi_{b}^{\prime*0}K^{*-}$  & $7.85\times10^{-20}$  & $9.55\times10^{-8}$  & $\Omega_{bb}^{-}\to\Xi_{b}^{\prime*0}D^{-}$  & $6.82\times10^{-20}$  & $8.29\times10^{-8}$ \tabularnewline
		$\Omega_{bb}^{-}\to\Xi_{b}^{\prime*0}D^{*-}$  & $1.23\times10^{-19}$  & $1.49\times10^{-7}$  & $\Omega_{bb}^{-}\to\Xi_{b}^{\prime*0}D_{s}^{-}$  & $1.78\times10^{-18}$  & $2.17\times10^{-6}$ \tabularnewline
		$\Omega_{bb}^{-}\to\Xi_{b}^{\prime*0}D_{s}^{*-}$  & $3.02\times10^{-18}$  & $3.67\times10^{-6}$  &  &  & \tabularnewline
		\hline 
		$\Omega_{bb}^{-}\to\Omega_{bc}^{*0}\pi^{-}$  & $7.77\times10^{-16}$  & $9.45\times10^{-4}$  & $\Omega_{bb}^{-}\to\Omega_{bc}^{*0}\rho^{-}$  & $2.11\times10^{-15}$  & $2.57\times10^{-3}$ \tabularnewline
		$\Omega_{bb}^{-}\to\Omega_{bc}^{*0}a_{1}^{-}$  & $2.73\times10^{-15}$  & $3.32\times10^{-3}$  & $\Omega_{bb}^{-}\to\Omega_{bc}^{*0}K^{-}$  & $5.98\times10^{-17}$  & $7.27\times10^{-5}$ \tabularnewline
		$\Omega_{bb}^{-}\to\Omega_{bc}^{*0}K^{*-}$  & $1.08\times10^{-16}$  & $1.32\times10^{-4}$  & $\Omega_{bb}^{-}\to\Omega_{bc}^{*0}D^{-}$  & $4.76\times10^{-17}$  & $5.79\times10^{-5}$ \tabularnewline
		$\Omega_{bb}^{-}\to\Omega_{bc}^{*0}D^{*-}$  & $1.34\times10^{-16}$  & $1.62\times10^{-4}$  & $\Omega_{bb}^{-}\to\Omega_{bc}^{*0}D_{s}^{-}$  & $1.14\times10^{-15}$  & $1.39\times10^{-3}$ \tabularnewline
		$\Omega_{bb}^{-}\to\Omega_{bc}^{*0}D_{s}^{*-}$  & $3.15\times10^{-15}$  & $3.83\times10^{-3}$  &  &  & \tabularnewline
		\hline 
	\end{tabular}
\end{table}

\begin{table}
	\caption{Nonleptonic decays for $bc$ sector with the $c$ quark decay.}
	\label{Tab:non_bc_c}%
	\begin{tabular}{l|c|c|l|c|c}
		\hline 
		channels  & $\Gamma/\text{~GeV}$  & ${\cal B}$  & channels  & $\Gamma/\text{~GeV}$ & ${\cal B}$ \tabularnewline
		\hline 
		$\Xi_{bc}^{+}\to\Sigma_{b}^{*0}\pi^{+}$  & $3.80\times10^{-16}$  & $1.41\times10^{-4}$  & $\Xi_{bc}^{+}\to\Sigma_{b}^{*0}\rho^{+}$  & $3.24\times10^{-15}$  & $1.20\times10^{-3}$ \tabularnewline
		$\Xi_{bc}^{+}\to\Sigma_{b}^{*0}K^{*+}$  & $1.77\times10^{-16}$  & $6.57\times10^{-5}$  & $\Xi_{bc}^{+}\to\Sigma_{b}^{*0}K^{+}$  & $2.32\times10^{-17}$  & $8.60\times10^{-6}$ \tabularnewline
		\hline 
		$\Xi_{bc}^{+}\to\Xi_{b}^{\prime*0}\pi^{+}$  & $7.63\times10^{-15}$  & $2.83\times10^{-3}$  & $\Xi_{bc}^{+}\to\Xi_{b}^{\prime*0}\rho^{+}$  & $5.06\times10^{-14}$  & $1.88\times10^{-2}$ \tabularnewline
		$\Xi_{bc}^{+}\to\Xi_{b}^{\prime*0}K^{*+}$  & $2.20\times10^{-15}$  & $8.18\times10^{-4}$  & $\Xi_{bc}^{+}\to\Xi_{b}^{\prime*0}K^{+}$  & $3.69\times10^{-16}$  & $1.37\times10^{-4}$ \tabularnewline
		\hline 
		$\Xi_{bc}^{0}\to\Sigma_{b}^{*-}\pi^{+}$  & $7.60\times10^{-16}$  & $1.07\times10^{-4}$  & $\Xi_{bc}^{0}\to\Sigma_{b}^{*-}\rho^{+}$  & $6.43\times10^{-15}$  & $9.09\times10^{-4}$ \tabularnewline
		$\Xi_{bc}^{0}\to\Sigma_{b}^{*-}K^{*+}$  & $3.50\times10^{-16}$  & $4.95\times10^{-5}$  & $\Xi_{bc}^{0}\to\Sigma_{b}^{*-}K^{+}$  & $4.62\times10^{-17}$  & $6.53\times10^{-6}$ \tabularnewline
		\hline 
		$\Xi_{bc}^{0}\to\Xi_{b}^{\prime*-}\pi^{+}$  & $7.67\times10^{-15}$  & $1.08\times10^{-3}$  & $\Xi_{bc}^{0}\to\Xi_{b}^{\prime*-}\rho^{+}$  & $4.85\times10^{-14}$  & $6.86\times10^{-3}$ \tabularnewline
		$\Xi_{bc}^{0}\to\Xi_{b}^{\prime*-}K^{*+}$  & $2.05\times10^{-15}$  & $2.90\times10^{-4}$  & $\Xi_{bc}^{0}\to\Xi_{b}^{\prime*-}K^{+}$  & $3.62\times10^{-16}$  & $5.11\times10^{-5}$ \tabularnewline
		\hline 
		$\Omega_{bc}^{0}\to\Xi_{b}^{\prime*-}\pi^{+}$  & $4.12\times10^{-16}$  & $1.38\times10^{-4}$  & $\Omega_{bc}^{0}\to\Xi_{b}^{\prime*-}\rho^{+}$  & $2.38\times10^{-15}$  & $7.97\times10^{-4}$ \tabularnewline
		$\Omega_{bc}^{0}\to\Xi_{b}^{\prime*-}K^{*+}$  & $1.10\times10^{-16}$  & $3.67\times10^{-5}$  & $\Omega_{bc}^{0}\to\Xi_{b}^{\prime*-}K^{+}$  & $2.07\times10^{-17}$  & $6.91\times10^{-6}$ \tabularnewline
		\hline 
		$\Omega_{bc}^{0}\to\Omega_{b}^{*-}\pi^{+}$  & $1.68\times10^{-14}$  & $5.61\times10^{-3}$  & $\Omega_{bc}^{0}\to\Omega_{b}^{*-}\rho^{+}$  & $4.85\times10^{-14}$  & $1.62\times10^{-2}$ \tabularnewline
		$\Omega_{bc}^{0}\to\Omega_{b}^{*-}K^{*+}$  & $9.07\times10^{-16}$  & $3.03\times10^{-4}$  & $\Omega_{bc}^{0}\to\Omega_{b}^{*-}K^{+}$  & $5.48\times10^{-16}$  & $1.83\times10^{-4}$ \tabularnewline
		\hline 
	\end{tabular}
\end{table}

\begin{table}
	\caption{Nonleptonic decays for $bc$ sector with the $b$ quark decay.}
	\label{Tab:non_bc_b}%
	\begin{tabular}{l|c|c|l|c|c}
		\hline 
		channels  & $\Gamma/\text{~GeV}$  & ${\cal B}$  & channels  & $\Gamma/\text{~GeV}$ & ${\cal B}$ \tabularnewline
		\hline 
		$\Xi_{bc}^{+}\to\Sigma_{c}^{*++}\pi^{-}$  & $8.19\times10^{-19}$  & $3.04\times10^{-7}$  & $\Xi_{bc}^{+}\to\Sigma_{c}^{*++}\rho^{-}$  & $1.85\times10^{-18}$  & $6.87\times10^{-7}$ \tabularnewline
		$\Xi_{bc}^{+}\to\Sigma_{c}^{*++}a_{1}^{-}$  & $2.63\times10^{-18}$  & $9.76\times10^{-7}$  & $\Xi_{bc}^{+}\to\Sigma_{c}^{*++}K^{-}$  & $6.60\times10^{-20}$  & $2.45\times10^{-8}$ \tabularnewline
		$\Xi_{bc}^{+}\to\Sigma_{c}^{*++}K^{*-}$  & $9.68\times10^{-20}$  & $3.59\times10^{-8}$  & $\Xi_{bc}^{+}\to\Sigma_{c}^{*++}D^{-}$  & $1.14\times10^{-19}$  & $4.22\times10^{-8}$ \tabularnewline
		$\Xi_{bc}^{+}\to\Sigma_{c}^{*++}D^{*-}$  & $1.70\times10^{-19}$  & $6.32\times10^{-8}$  & $\Xi_{bc}^{+}\to\Sigma_{c}^{*++}D_{s}^{-}$  & $3.03\times10^{-18}$  & $1.12\times10^{-6}$ \tabularnewline
		$\Xi_{bc}^{+}\to\Sigma_{c}^{*++}D_{s}^{*-}$  & $4.23\times10^{-18}$  & $1.57\times10^{-6}$  &  &  & \tabularnewline
		\hline 
		$\Xi_{bc}^{+}\to\Xi_{cc}^{*++}\pi^{-}$  & $8.44\times10^{-16}$  & $3.13\times10^{-4}$  & $\Xi_{bc}^{+}\to\Xi_{cc}^{*++}\rho^{-}$  & $2.21\times10^{-15}$  & $8.20\times10^{-4}$ \tabularnewline
		$\Xi_{bc}^{+}\to\Xi_{cc}^{*++}a_{1}^{-}$  & $2.98\times10^{-15}$  & $1.10\times10^{-3}$  & $\Xi_{bc}^{+}\to\Xi_{cc}^{*++}K^{-}$  & $6.59\times10^{-17}$  & $2.44\times10^{-5}$ \tabularnewline
		$\Xi_{bc}^{+}\to\Xi_{cc}^{*++}K^{*-}$  & $1.14\times10^{-16}$  & $4.24\times10^{-5}$  & $\Xi_{bc}^{+}\to\Xi_{cc}^{*++}D^{-}$  & $6.55\times10^{-17}$  & $2.43\times10^{-5}$ \tabularnewline
		$\Xi_{bc}^{+}\to\Xi_{cc}^{*++}D^{*-}$  & $1.61\times10^{-16}$  & $5.98\times10^{-5}$  & $\Xi_{bc}^{+}\to\Xi_{cc}^{*++}D_{s}^{-}$  & $1.62\times10^{-15}$  & $6.00\times10^{-4}$ \tabularnewline
		$\Xi_{bc}^{+}\to\Xi_{cc}^{*++}D_{s}^{*-}$  & $3.87\times10^{-15}$  & $1.44\times10^{-3}$  &  &  & \tabularnewline
		\hline 
		$\Xi_{bc}^{0}\to\Sigma_{c}^{*+}\pi^{-}$  & $4.09\times10^{-19}$  & $5.79\times10^{-8}$  & $\Xi_{bc}^{0}\to\Sigma_{c}^{*+}\rho^{-}$  & $9.26\times10^{-19}$  & $1.31\times10^{-7}$ \tabularnewline
		$\Xi_{bc}^{0}\to\Sigma_{c}^{*+}a_{1}^{-}$  & $1.32\times10^{-18}$  & $1.86\times10^{-7}$  & $\Xi_{bc}^{0}\to\Sigma_{c}^{*+}K^{-}$  & $3.30\times10^{-20}$  & $4.66\times10^{-9}$ \tabularnewline
		$\Xi_{bc}^{0}\to\Sigma_{c}^{*+}K^{*-}$  & $4.84\times10^{-20}$  & $6.84\times10^{-9}$  & $\Xi_{bc}^{0}\to\Sigma_{c}^{*+}D^{-}$  & $5.69\times10^{-20}$  & $8.05\times10^{-9}$ \tabularnewline
		$\Xi_{bc}^{0}\to\Sigma_{c}^{*+}D^{*-}$  & $8.52\times10^{-20}$  & $1.20\times10^{-8}$  & $\Xi_{bc}^{0}\to\Sigma_{c}^{*+}D_{s}^{-}$  & $1.52\times10^{-18}$  & $2.14\times10^{-7}$ \tabularnewline
		$\Xi_{bc}^{0}\to\Sigma_{c}^{*+}D_{s}^{*-}$  & $2.12\times10^{-18}$  & $2.99\times10^{-7}$  &  &  & \tabularnewline
		\hline 
		$\Xi_{bc}^{0}\to\Xi_{cc}^{*+}\pi^{-}$  & $8.44\times10^{-16}$  & $1.19\times10^{-4}$  & $\Xi_{bc}^{0}\to\Xi_{cc}^{*+}\rho^{-}$  & $2.21\times10^{-15}$  & $3.13\times10^{-4}$ \tabularnewline
		$\Xi_{bc}^{0}\to\Xi_{cc}^{*+}a_{1}^{-}$  & $2.98\times10^{-15}$  & $4.21\times10^{-4}$  & $\Xi_{bc}^{0}\to\Xi_{cc}^{*+}K^{-}$  & $6.59\times10^{-17}$  & $9.31\times10^{-6}$ \tabularnewline
		$\Xi_{bc}^{0}\to\Xi_{cc}^{*+}K^{*-}$  & $1.14\times10^{-16}$  & $1.62\times10^{-5}$  & $\Xi_{bc}^{0}\to\Xi_{cc}^{*+}D^{-}$  & $6.55\times10^{-17}$  & $9.26\times10^{-6}$ \tabularnewline
		$\Xi_{bc}^{0}\to\Xi_{cc}^{*+}D^{*-}$  & $1.61\times10^{-16}$  & $2.28\times10^{-5}$  & $\Xi_{bc}^{0}\to\Xi_{cc}^{*+}D_{s}^{-}$  & $1.62\times10^{-15}$  & $2.29\times10^{-4}$ \tabularnewline
		$\Xi_{bc}^{0}\to\Xi_{cc}^{*+}D_{s}^{*-}$  & $3.87\times10^{-15}$  & $5.47\times10^{-4}$  &  &  & \tabularnewline
		\hline 
		$\Omega_{bc}^{0}\to\Xi_{c}^{\prime*+}\pi^{-}$  & $3.54\times10^{-19}$  & $1.19\times10^{-7}$  & $\Omega_{bc}^{0}\to\Xi_{c}^{\prime*+}\rho^{-}$  & $8.10\times10^{-19}$  & $2.71\times10^{-7}$ \tabularnewline
		$\Omega_{bc}^{0}\to\Xi_{c}^{\prime*+}a_{1}^{-}$  & $1.14\times10^{-18}$  & $3.81\times10^{-7}$  & $\Omega_{bc}^{0}\to\Xi_{c}^{\prime*+}K^{-}$  & $2.85\times10^{-20}$  & $9.53\times10^{-9}$ \tabularnewline
		$\Omega_{bc}^{0}\to\Xi_{c}^{\prime*+}K^{*-}$  & $4.22\times10^{-20}$  & $1.41\times10^{-8}$  & $\Omega_{bc}^{0}\to\Xi_{c}^{\prime*+}D^{-}$  & $4.76\times10^{-20}$  & $1.59\times10^{-8}$ \tabularnewline
		$\Omega_{bc}^{0}\to\Xi_{c}^{\prime*+}D^{*-}$  & $7.24\times10^{-20}$  & $2.42\times10^{-8}$  & $\Omega_{bc}^{0}\to\Xi_{c}^{\prime*+}D_{s}^{-}$  & $1.26\times10^{-18}$  & $4.22\times10^{-7}$ \tabularnewline
		$\Omega_{bc}^{0}\to\Xi_{c}^{\prime*+}D_{s}^{*-}$  & $1.79\times10^{-18}$  & $6.00\times10^{-7}$  &  &  & \tabularnewline
		\hline 
		$\Omega_{bc}^{0}\to\Omega_{cc}^{*+}\pi^{-}$  & $8.33\times10^{-16}$  & $2.79\times10^{-4}$  & $\Omega_{bc}^{0}\to\Omega_{cc}^{*+}\rho^{-}$  & $2.16\times10^{-15}$  & $7.22\times10^{-4}$ \tabularnewline
		$\Omega_{bc}^{0}\to\Omega_{cc}^{*+}a_{1}^{-}$  & $2.83\times10^{-15}$  & $9.46\times10^{-4}$  & $\Omega_{bc}^{0}\to\Omega_{cc}^{*+}K^{-}$  & $6.46\times10^{-17}$  & $2.16\times10^{-5}$ \tabularnewline
		$\Omega_{bc}^{0}\to\Omega_{cc}^{*+}K^{*-}$  & $1.11\times10^{-16}$  & $3.71\times10^{-5}$  & $\Omega_{bc}^{0}\to\Omega_{cc}^{*+}D^{-}$  & $5.81\times10^{-17}$  & $1.94\times10^{-5}$ \tabularnewline
		$\Omega_{bc}^{0}\to\Omega_{cc}^{*+}D^{*-}$  & $1.44\times10^{-16}$  & $4.82\times10^{-5}$  & $\Omega_{bc}^{0}\to\Omega_{cc}^{*+}D_{s}^{-}$  & $1.41\times10^{-15}$  & $4.72\times10^{-4}$ \tabularnewline
		$\Omega_{bc}^{0}\to\Omega_{cc}^{*+}D_{s}^{*-}$  & $3.43\times10^{-15}$  & $1.15\times10^{-3}$  &  &  & \tabularnewline
		\hline 
	\end{tabular}
\end{table}

\begin{table}
	\caption{Nonleptonic decays for $bc^{\prime}$ sector with the $c$
		quark decay.}
	\label{Tab:non_bcp_c}%
	\begin{tabular}{l|c|c|l|c|c}
		\hline 
		channels  & $\Gamma/\text{~GeV}$  & ${\cal B}$  & channels  & $\Gamma/\text{~GeV}$ & ${\cal B}$ \tabularnewline
		\hline 
		$\Xi_{bc}^{\prime+}\to\Sigma_{b}^{*0}\pi^{+}$  & $1.14\times10^{-15}$  & $4.22\times10^{-4}$  & $\Xi_{bc}^{\prime+}\to\Sigma_{b}^{*0}\rho^{+}$  & $9.73\times10^{-15}$  & $3.61\times10^{-3}$ \tabularnewline
		$\Xi_{bc}^{\prime+}\to\Sigma_{b}^{*0}K^{*+}$  & $5.32\times10^{-16}$  & $1.97\times10^{-4}$  & $\Xi_{bc}^{\prime+}\to\Sigma_{b}^{*0}K^{+}$  & $6.95\times10^{-17}$  & $2.58\times10^{-5}$ \tabularnewline
		\hline 
		$\Xi_{bc}^{\prime+}\to\Xi_{b}^{\prime*0}\pi^{+}$  & $2.29\times10^{-14}$  & $8.49\times10^{-3}$  & $\Xi_{bc}^{\prime+}\to\Xi_{b}^{\prime*0}\rho^{+}$  & $1.52\times10^{-13}$  & $5.63\times10^{-2}$ \tabularnewline
		$\Xi_{bc}^{\prime+}\to\Xi_{b}^{\prime*0}K^{*+}$  & $6.61\times10^{-15}$  & $2.45\times10^{-3}$  & $\Xi_{bc}^{\prime+}\to\Xi_{b}^{\prime*0}K^{+}$  & $1.11\times10^{-15}$  & $4.10\times10^{-4}$ \tabularnewline
		\hline 
		$\Xi_{bc}^{\prime0}\to\Sigma_{b}^{*-}\pi^{+}$  & $2.28\times10^{-15}$  & $3.22\times10^{-4}$  & $\Xi_{bc}^{\prime0}\to\Sigma_{b}^{*-}\rho^{+}$  & $1.93\times10^{-14}$  & $2.73\times10^{-3}$ \tabularnewline
		$\Xi_{bc}^{\prime0}\to\Sigma_{b}^{*-}K^{*+}$  & $1.05\times10^{-15}$  & $1.48\times10^{-4}$  & $\Xi_{bc}^{\prime0}\to\Sigma_{b}^{*-}K^{+}$  & $1.39\times10^{-16}$  & $1.96\times10^{-5}$ \tabularnewline
		\hline 
		$\Xi_{bc}^{\prime0}\to\Xi_{b}^{\prime*-}\pi^{+}$  & $2.30\times10^{-14}$  & $3.25\times10^{-3}$  & $\Xi_{bc}^{\prime0}\to\Xi_{b}^{\prime*-}\rho^{+}$  & $1.46\times10^{-13}$  & $2.06\times10^{-2}$ \tabularnewline
		$\Xi_{bc}^{\prime0}\to\Xi_{b}^{\prime*-}K^{*+}$  & $6.16\times10^{-15}$  & $8.70\times10^{-4}$  & $\Xi_{bc}^{\prime0}\to\Xi_{b}^{\prime*-}K^{+}$  & $1.08\times10^{-15}$  & $1.53\times10^{-4}$ \tabularnewline
		\hline 
		$\Omega_{bc}^{\prime0}\to\Xi_{b}^{\prime*-}\pi^{+}$  & $1.23\times10^{-15}$  & $4.13\times10^{-4}$  & $\Omega_{bc}^{\prime0}\to\Xi_{b}^{\prime*-}\rho^{+}$  & $7.15\times10^{-15}$  & $2.39\times10^{-3}$ \tabularnewline
		$\Omega_{bc}^{\prime0}\to\Xi_{b}^{\prime*-}K^{*+}$  & $3.29\times10^{-16}$  & $1.10\times10^{-4}$  & $\Omega_{bc}^{\prime0}\to\Xi_{b}^{\prime*-}K^{+}$  & $6.20\times10^{-17}$  & $2.07\times10^{-5}$ \tabularnewline
		\hline 
		$\Omega_{bc}^{\prime0}\to\Omega_{b}^{*-}\pi^{+}$  & $5.03\times10^{-14}$  & $1.68\times10^{-2}$  & $\Omega_{bc}^{\prime0}\to\Omega_{b}^{*-}\rho^{+}$  & $1.46\times10^{-13}$  & $4.87\times10^{-2}$ \tabularnewline
		$\Omega_{bc}^{\prime0}\to\Omega_{b}^{*-}K^{*+}$  & $2.72\times10^{-15}$  & $9.10\times10^{-4}$  & $\Omega_{bc}^{\prime0}\to\Omega_{b}^{*-}K^{+}$  & $1.65\times10^{-15}$  & $5.50\times10^{-4}$ \tabularnewline
		\hline 
	\end{tabular}
\end{table}

\begin{table}
	\caption{Nonleptonic decays for $bc^{\prime}$ sector with the $b$
		quark decay.}
	\label{Tab:non_bcp_b}%
	\begin{tabular}{l|c|c|l|c|c}
		\hline 
		channels  & $\Gamma/\text{~GeV}$  & ${\cal B}$  & channels  & $\Gamma/\text{~GeV}$ & ${\cal B}$ \tabularnewline
		\hline 
		$\Xi_{bc}^{\prime+}\to\Sigma_{c}^{*++}\pi^{-}$  & $2.46\times10^{-18}$  & $9.11\times10^{-7}$  & $\Xi_{bc}^{\prime+}\to\Sigma_{c}^{*++}\rho^{-}$  & $5.55\times10^{-18}$  & $2.06\times10^{-6}$ \tabularnewline
		$\Xi_{bc}^{\prime+}\to\Sigma_{c}^{*++}a_{1}^{-}$  & $7.90\times10^{-18}$  & $2.93\times10^{-6}$  & $\Xi_{bc}^{\prime+}\to\Sigma_{c}^{*++}K^{-}$  & $1.98\times10^{-19}$  & $7.34\times10^{-8}$ \tabularnewline
		$\Xi_{bc}^{\prime+}\to\Sigma_{c}^{*++}K^{*-}$  & $2.90\times10^{-19}$  & $1.08\times10^{-7}$  & $\Xi_{bc}^{\prime+}\to\Sigma_{c}^{*++}D^{-}$  & $3.42\times10^{-19}$  & $1.27\times10^{-7}$ \tabularnewline
		$\Xi_{bc}^{\prime+}\to\Sigma_{c}^{*++}D^{*-}$  & $5.11\times10^{-19}$  & $1.90\times10^{-7}$  & $\Xi_{bc}^{\prime+}\to\Sigma_{c}^{*++}D_{s}^{-}$  & $9.10\times10^{-18}$  & $3.37\times10^{-6}$ \tabularnewline
		$\Xi_{bc}^{\prime+}\to\Sigma_{c}^{*++}D_{s}^{*-}$  & $1.27\times10^{-17}$  & $4.71\times10^{-6}$  &  &  & \tabularnewline
		\hline 
		$\Xi_{bc}^{\prime+}\to\Xi_{cc}^{*++}\pi^{-}$  & $2.53\times10^{-15}$  & $9.39\times10^{-4}$  & $\Xi_{bc}^{\prime+}\to\Xi_{cc}^{*++}\rho^{-}$  & $6.63\times10^{-15}$  & $2.46\times10^{-3}$ \tabularnewline
		$\Xi_{bc}^{\prime+}\to\Xi_{cc}^{*++}a_{1}^{-}$  & $8.93\times10^{-15}$  & $3.31\times10^{-3}$  & $\Xi_{bc}^{\prime+}\to\Xi_{cc}^{*++}K^{-}$  & $1.98\times10^{-16}$  & $7.33\times10^{-5}$ \tabularnewline
		$\Xi_{bc}^{\prime+}\to\Xi_{cc}^{*++}K^{*-}$  & $3.43\times10^{-16}$  & $1.27\times10^{-4}$  & $\Xi_{bc}^{\prime+}\to\Xi_{cc}^{*++}D^{-}$  & $1.97\times10^{-16}$  & $7.29\times10^{-5}$ \tabularnewline
		$\Xi_{bc}^{\prime+}\to\Xi_{cc}^{*++}D^{*-}$  & $4.84\times10^{-16}$  & $1.79\times10^{-4}$  & $\Xi_{bc}^{\prime+}\to\Xi_{cc}^{*++}D_{s}^{-}$  & $4.85\times10^{-15}$  & $1.80\times10^{-3}$ \tabularnewline
		$\Xi_{bc}^{\prime+}\to\Xi_{cc}^{*++}D_{s}^{*-}$  & $1.16\times10^{-14}$  & $4.31\times10^{-3}$  &  &  & \tabularnewline
		\hline 
		$\Xi_{bc}^{\prime0}\to\Sigma_{c}^{*+}\pi^{-}$  & $1.23\times10^{-18}$  & $1.74\times10^{-7}$  & $\Xi_{bc}^{\prime0}\to\Sigma_{c}^{*+}\rho^{-}$  & $2.78\times10^{-18}$  & $3.93\times10^{-7}$ \tabularnewline
		$\Xi_{bc}^{\prime0}\to\Sigma_{c}^{*+}a_{1}^{-}$  & $3.95\times10^{-18}$  & $5.58\times10^{-7}$  & $\Xi_{bc}^{\prime0}\to\Sigma_{c}^{*+}K^{-}$  & $9.90\times10^{-20}$  & $1.40\times10^{-8}$ \tabularnewline
		$\Xi_{bc}^{\prime0}\to\Sigma_{c}^{*+}K^{*-}$  & $1.45\times10^{-19}$  & $2.05\times10^{-8}$  & $\Xi_{bc}^{\prime0}\to\Sigma_{c}^{*+}D^{-}$  & $1.71\times10^{-19}$  & $2.42\times10^{-8}$ \tabularnewline
		$\Xi_{bc}^{\prime0}\to\Sigma_{c}^{*+}D^{*-}$  & $2.55\times10^{-19}$  & $3.61\times10^{-8}$  & $\Xi_{bc}^{\prime0}\to\Sigma_{c}^{*+}D_{s}^{-}$  & $4.55\times10^{-18}$  & $6.43\times10^{-7}$ \tabularnewline
		$\Xi_{bc}^{\prime0}\to\Sigma_{c}^{*+}D_{s}^{*-}$  & $6.35\times10^{-18}$  & $8.98\times10^{-7}$  &  &  & \tabularnewline
		\hline 
		$\Xi_{bc}^{\prime0}\to\Xi_{cc}^{*+}\pi^{-}$  & $2.53\times10^{-15}$  & $3.58\times10^{-4}$  & $\Xi_{bc}^{\prime0}\to\Xi_{cc}^{*+}\rho^{-}$  & $6.63\times10^{-15}$  & $9.38\times10^{-4}$ \tabularnewline
		$\Xi_{bc}^{\prime0}\to\Xi_{cc}^{*+}a_{1}^{-}$  & $8.93\times10^{-15}$  & $1.26\times10^{-3}$  & $\Xi_{bc}^{\prime0}\to\Xi_{cc}^{*+}K^{-}$  & $1.98\times10^{-16}$  & $2.79\times10^{-5}$ \tabularnewline
		$\Xi_{bc}^{\prime0}\to\Xi_{cc}^{*+}K^{*-}$  & $3.43\times10^{-16}$  & $4.85\times10^{-5}$  & $\Xi_{bc}^{\prime0}\to\Xi_{cc}^{*+}D^{-}$  & $1.97\times10^{-16}$  & $2.78\times10^{-5}$ \tabularnewline
		$\Xi_{bc}^{\prime0}\to\Xi_{cc}^{*+}D^{*-}$  & $4.84\times10^{-16}$  & $6.84\times10^{-5}$  & $\Xi_{bc}^{\prime0}\to\Xi_{cc}^{*+}D_{s}^{-}$  & $4.85\times10^{-15}$  & $6.86\times10^{-4}$ \tabularnewline
		$\Xi_{bc}^{\prime0}\to\Xi_{cc}^{*+}D_{s}^{*-}$  & $1.16\times10^{-14}$  & $1.64\times10^{-3}$  &  &  & \tabularnewline
		\hline 
		$\Omega_{bc}^{\prime0}\to\Xi_{c}^{\prime*+}\pi^{-}$  & $1.06\times10^{-18}$  & $3.56\times10^{-7}$  & $\Omega_{bc}^{\prime0}\to\Xi_{c}^{\prime*+}\rho^{-}$  & $2.43\times10^{-18}$  & $8.12\times10^{-7}$ \tabularnewline
		$\Omega_{bc}^{\prime0}\to\Xi_{c}^{\prime*+}a_{1}^{-}$  & $3.42\times10^{-18}$  & $1.14\times10^{-6}$  & $\Omega_{bc}^{\prime0}\to\Xi_{c}^{\prime*+}K^{-}$  & $8.55\times10^{-20}$  & $2.86\times10^{-8}$ \tabularnewline
		$\Omega_{bc}^{\prime0}\to\Xi_{c}^{\prime*+}K^{*-}$  & $1.27\times10^{-19}$  & $4.24\times10^{-8}$  & $\Omega_{bc}^{\prime0}\to\Xi_{c}^{\prime*+}D^{-}$  & $1.43\times10^{-19}$  & $4.78\times10^{-8}$ \tabularnewline
		$\Omega_{bc}^{\prime0}\to\Xi_{c}^{\prime*+}D^{*-}$  & $2.17\times10^{-19}$  & $7.26\times10^{-8}$  & $\Omega_{bc}^{\prime0}\to\Xi_{c}^{\prime*+}D_{s}^{-}$  & $3.79\times10^{-18}$  & $1.27\times10^{-6}$ \tabularnewline
		$\Omega_{bc}^{\prime0}\to\Xi_{c}^{\prime*+}D_{s}^{*-}$  & $5.38\times10^{-18}$  & $1.80\times10^{-6}$  &  &  & \tabularnewline
		\hline 
		$\Omega_{bc}^{\prime0}\to\Omega_{cc}^{*+}\pi^{-}$  & $2.50\times10^{-15}$  & $8.36\times10^{-4}$  & $\Omega_{bc}^{\prime0}\to\Omega_{cc}^{*+}\rho^{-}$  & $6.47\times10^{-15}$  & $2.16\times10^{-3}$ \tabularnewline
		$\Omega_{bc}^{\prime0}\to\Omega_{cc}^{*+}a_{1}^{-}$  & $8.49\times10^{-15}$  & $2.84\times10^{-3}$  & $\Omega_{bc}^{\prime0}\to\Omega_{cc}^{*+}K^{-}$  & $1.94\times10^{-16}$  & $6.48\times10^{-5}$ \tabularnewline
		$\Omega_{bc}^{\prime0}\to\Omega_{cc}^{*+}K^{*-}$  & $3.33\times10^{-16}$  & $1.11\times10^{-4}$  & $\Omega_{bc}^{\prime0}\to\Omega_{cc}^{*+}D^{-}$  & $1.74\times10^{-16}$  & $5.83\times10^{-5}$ \tabularnewline
		$\Omega_{bc}^{\prime0}\to\Omega_{cc}^{*+}D^{*-}$  & $4.33\times10^{-16}$  & $1.45\times10^{-4}$  & $\Omega_{bc}^{\prime0}\to\Omega_{cc}^{*+}D_{s}^{-}$  & $4.24\times10^{-15}$  & $1.42\times10^{-3}$ \tabularnewline
		$\Omega_{bc}^{\prime0}\to\Omega_{cc}^{*+}D_{s}^{*-}$  & $1.03\times10^{-14}$  & $3.44\times10^{-3}$  &  &  & \tabularnewline
		\hline 
	\end{tabular}
\end{table}

\subsection{SU(3) symmetry for semi-leptonic decays}

According to  the flavor SU(3) symmetry, there exist the following relations
among these semileptonic decay widths \cite{Wang:2017azm}, which can also be readily rederived using the overlapping factors given in Table \ref{Tab:overlapping_factors}:
\begin{itemize}
	\item $cc$ sector
	\begin{eqnarray}
		&  & \frac{\Gamma(\Xi_{cc}^{++}\to\Sigma_{c}^{*+}l^{+}\nu)}{|V_{cd}|^{2}}=\frac{\Gamma(\Xi_{cc}^{+}\to\Sigma_{c}^{*0}l^{+}\nu)}{2|V_{cd}|^{2}}=\frac{\Gamma(\Omega_{cc}^{+}\to\Xi_{c}^{\prime*0}l^{+}\nu)}{|V_{cd}|^{2}}\nonumber \\
		& = & \frac{\Gamma(\Xi_{cc}^{++}\to\Xi_{c}^{\prime*+}l^{+}\nu)}{|V_{cs}|^{2}}=\frac{\Gamma(\Xi_{cc}^{+}\to\Xi_{c}^{\prime*0}l^{+}\nu)}{|V_{cs}|^{2}}=\frac{\Gamma(\Omega_{cc}^{+}\to\Omega_{c}^{*0}l^{+}\nu)}{2|V_{cs}|^{2}},
	\end{eqnarray}
	\item $bb$ sector
	\begin{eqnarray}
		&  & \Gamma(\Xi_{bb}^{0}\to\Sigma_{b}^{*+}l^{-}\bar{\nu})=2\Gamma(\Xi_{bb}^{-}\to\Sigma_{b}^{*0}l^{-}\bar{\nu})=2\Gamma(\Omega_{bb}^{-}\to\Xi_{b}^{\prime*0}l^{-}\bar{\nu}),\nonumber \\
		&  & \Gamma(\Xi_{bb}^{0}\to\Xi_{bc}^{*+}l^{-}\bar{\nu})=\Gamma(\Xi_{bb}^{-}\to\Xi_{bc}^{*0}l^{-}\bar{\nu})=\Gamma(\Omega_{bb}^{-}\to\Omega_{bc}^{*0}l^{-}\bar{\nu}),
	\end{eqnarray}
	\item $bc$ sector with the $c$ quark decay
	\begin{eqnarray}
		&  & \frac{\Gamma(\Xi_{bc}^{+}\to\Sigma_{b}^{*0}l^{+}\nu)}{|V_{cd}|^{2}}=\frac{\Gamma(\Xi_{bc}^{0}\to\Sigma_{b}^{*-}l^{+}\nu)}{2|V_{cd}|^{2}}=\frac{\Gamma(\Omega_{bc}^{0}\to\Xi_{b}^{\prime*-}l^{+}\nu)}{|V_{cd}|^{2}}\nonumber \\
		& = & \frac{\Gamma(\Xi_{bc}^{+}\to\Xi_{b}^{\prime*0}l^{+}\nu)}{|V_{cs}|^{2}}=\frac{\Gamma(\Xi_{bc}^{0}\to\Xi_{b}^{\prime*-}l^{+}\nu)}{|V_{cs}|^{2}}=\frac{\Gamma(\Omega_{bc}^{0}\to\Omega_{b}^{*-}l^{+}\nu)}{2|V_{cs}|^{2}},
	\end{eqnarray}
	\item $bc$ sector with the $b$ quark decay
	\begin{eqnarray}
		&  & \Gamma(\Xi_{bc}^{+}\to\Sigma_{c}^{*++}l^{-}\bar{\nu})=2\Gamma(\Xi_{bc}^{0}\to\Sigma_{c}^{*+}l^{-}\bar{\nu})=2\Gamma(\Omega_{bc}^{0}\to\Xi_{c}^{\prime*+}l^{-}\bar{\nu}),\nonumber \\
		&  & \Gamma(\Xi_{bc}^{+}\to\Xi_{cc}^{*++}l^{-}\bar{\nu})=\Gamma(\Xi_{bc}^{0}\to\Xi_{cc}^{*+}l^{-}\bar{\nu})=\Gamma(\Omega_{bc}^{0}\to\Omega_{cc}^{*+}l^{-}\bar{\nu}),
	\end{eqnarray}
	\item $bc^{\prime}$ sector with the $c$ quark decay
	\begin{eqnarray}
		&  & \frac{\Gamma(\Xi_{bc}^{\prime+}\to\Sigma_{b}^{*0}l^{+}\nu)}{|V_{cd}|^{2}}=\frac{\Gamma(\Xi_{bc}^{\prime0}\to\Sigma_{b}^{*-}l^{+}\nu)}{2|V_{cd}|^{2}}=\frac{\Gamma(\Omega_{bc}^{\prime0}\to\Xi_{b}^{\prime*-}l^{+}\nu)}{|V_{cd}|^{2}}\nonumber \\
		& = & \frac{\Gamma(\Xi_{bc}^{\prime+}\to\Xi_{b}^{\prime*0}l^{+}\nu)}{|V_{cs}|^{2}}=\frac{\Gamma(\Xi_{bc}^{\prime0}\to\Xi_{b}^{\prime*-}l^{+}\nu)}{|V_{cs}|^{2}}=\frac{\Gamma(\Omega_{bc}^{\prime0}\to\Omega_{b}^{*-}l^{+}\nu)}{2|V_{cs}|^{2}},
	\end{eqnarray}
	\item $bc^{\prime}$ sector with the $b$ quark decay
	\begin{eqnarray}
		&  & \Gamma(\Xi_{bc}^{\prime+}\to\Sigma_{c}^{*++}l^{-}\bar{\nu})=2\Gamma(\Xi_{bc}^{\prime0}\to\Sigma_{c}^{*+}l^{-}\bar{\nu})=2\Gamma(\Omega_{bc}^{\prime0}\to\Xi_{c}^{\prime*+}l^{-}\bar{\nu}),\nonumber \\
		&  & \Gamma(\Xi_{bc}^{\prime+}\to\Xi_{cc}^{*++}l^{-}\bar{\nu})=\Gamma(\Xi_{bc}^{\prime0}\to\Xi_{cc}^{*+}l^{-}\bar{\nu})=\Gamma(\Omega_{bc}^{\prime0}\to\Omega_{cc}^{*+}l^{-}\bar{\nu}).
	\end{eqnarray}
\end{itemize}
Also, we have compared the predictions of the light-front approach
with those of SU(3) symmetry method taking $cc$ and $bb$ sectors as examples, which can be seen
in Tables \ref{Tab:SU3_breaking_cc} and \ref{Tab:SU3_breaking_bb}.

\begin{table}
	\caption{Quantitative predictions of SU(3) breaking for semi-leptonic decays:
		$cc$ sector. }
	\label{Tab:SU3_breaking_cc} %
	\begin{tabular}{l|c|c|c}
		\hline 
		channels  & $\Gamma/{\rm GeV}$ (LFQM)  & $\Gamma/{\rm GeV}$ (SU(3))  & $|{\rm LFQM}-{\rm SU(3)}|/{\rm SU(3)}$\tabularnewline
		\hline 
		$\Xi_{cc}^{++}\to\Sigma_{c}^{*+}e^{+}\nu_{e}$  & $1.26\times10^{-15}$  & $1.26\times10^{-15}$  & - -\tabularnewline
		$\Xi_{cc}^{+}\to\Sigma_{c}^{*0}e^{+}\nu_{e}$  & $2.51\times10^{-15}$  & $2.52\times10^{-15}$  & $0\%$\tabularnewline
		$\Omega_{cc}^{+}\to\Xi_{c}^{\prime*0}e^{+}\nu_{e}$  & $1.19\times10^{-15}$  & $1.26\times10^{-15}$  & $6\%$\tabularnewline
		$\Xi_{cc}^{++}\to\Xi_{c}^{\prime*+}e^{+}\nu_{e}$  & $1.61\times10^{-14}$  & $2.36\times10^{-14}$  & $32\%$\tabularnewline
		$\Xi_{cc}^{+}\to\Xi_{c}^{\prime*0}e^{+}\nu_{e}$  & $1.61\times10^{-14}$  & $2.36\times10^{-14}$  & $32\%$\tabularnewline
		$\Omega_{cc}^{+}\to\Omega_{c}^{*0}e^{+}\nu_{e}$  & $3.20\times10^{-14}$  & $4.72\times10^{-14}$  & $32\%$\tabularnewline
		\hline 
	\end{tabular}
\end{table}

\begin{table}
	\caption{Quantitative predictions of SU(3) breaking for semi-leptonic decays:
		$bb$ sector. }
	\label{Tab:SU3_breaking_bb} %
	\begin{tabular}{l|c|c|c}
		\hline 
		channels  & $\Gamma/{\rm GeV}$ (LFQM)  & $\Gamma/{\rm GeV}$ (SU(3))  & $|{\rm LFQM}-{\rm SU(3)}|/{\rm SU(3)}$\tabularnewline
		\hline 
		$\Xi_{bb}^{0}\to\Sigma_{b}^{*+}e^{-}\bar{\nu}_{e}$  & $3.88\times10^{-17}$  & $3.88\times10^{-17}$  & - -\tabularnewline
		$\Xi_{bb}^{-}\to\Sigma_{b}^{*0}e^{-}\bar{\nu}_{e}$  & $1.94\times10^{-17}$  & $1.94\times10^{-17}$  & $0\%$\tabularnewline
		$\Omega_{bb}^{-}\to\Xi_{b}^{\prime*0}e^{-}\bar{\nu}_{e}$  & $1.90\times10^{-17}$  & $1.94\times10^{-17}$  & $2\%$\tabularnewline
		\hline 
		$\Xi_{bb}^{0}\to\Xi_{bc}^{*+}e^{-}\bar{\nu}_{e}$  & $6.37\times10^{-15}$  & $6.37\times10^{-15}$  & - -\tabularnewline
		$\Xi_{bb}^{-}\to\Xi_{bc}^{*0}e^{-}\bar{\nu}_{e}$  & $6.37\times10^{-15}$  & $6.37\times10^{-15}$  & $0\%$\tabularnewline
		$\Omega_{bb}^{-}\to\Omega_{bc}^{*0}e^{-}\bar{\nu}_{e}$  & $7.03\times10^{-15}$  & $6.37\times10^{-15}$  & $10\%$\tabularnewline
		\hline 
	\end{tabular}
\end{table}

Some comments are given in order.
\begin{itemize}
	\item Note that, 1/2 to 3/2 process has completely the same SU(3) relations
	as the corresponding 1/2 to 1/2 case. This can be expected, because spin-3/2
	baryon shares the same flavor wave function as the corresponding spin-1/2 baryon.
	\item SU(3) predictions for the corresponding two channels in $bc$ and
	$bc^{\prime}$ sectors have completely the same form, as can be explained
	by the facts that they have the same final states and the formally fixed initial states as in Eqs. (\ref{eq:flavor_spin_bc})
	and (\ref{eq:flavor_spin_bcp}).
	\item We can see from Table \ref{Tab:SU3_breaking_cc} that, sizable SU(3)
	symmetry breaking takes place between the $c\to d$ and $c\to s$
	processes. Of course, this can be attributed to the model parameter:
	we have taken different quark mass for $d$ quark and $s$ quark. 
	\item The small deviation of 6\% in Table \ref{Tab:SU3_breaking_cc} and
	2\% in Table \ref{Tab:SU3_breaking_bb}, can be explained by the fact
	that we have taken a larger value for the mass of $Qs$ diquark than that of $Qu$ or
	$Qd$ diquark, where $Q=c/b$. And also, note that, SU(3) symmetry
	breaking in $c$ quark decay is usually larger than that in $b$ quark decay.
\end{itemize}

\subsection{Comparison}

\label{subsec:some_other_discussions}

For a comparison, we also list the results in Ref. \cite{Shi:2016gqt}. Some comments
will be given on the results of semi-leptonic and nonleptonic decays.
\begin{eqnarray}
{\cal B}(B_{c}\to B_{s}\bar{l}\nu) & = & 1.51\times10^{-2},\nonumber \\
{\cal B}(B_{c}\to B_{s}^{*}\bar{l}\nu) & = & 1.96\times10^{-2},\nonumber \\
{\cal B}(B_{c}\to B\bar{l}\nu) & = & 1.04\times10^{-3},\nonumber \\
{\cal B}(B_{c}\to B^{*}\bar{l}\nu) & = & 1.34\times10^{-3},\nonumber \\
{\cal B}(B_{c}\to B_{s}\pi) & = & 4.1\times10^{-2},\nonumber \\
{\cal B}(B_{c}\to B_{s}^{*}\pi) & = & 2.0\times10^{-2}.\label{eq:Bc_decay}
\end{eqnarray}

\begin{itemize}
	\item Since there exist large uncertainties in the lifetimes, we have also
	presented the results for decay widths.
	\item We find that the result for 1/2 to 3/2 process is roughly one order of magnitude
	smaller than the corresponding 1/2 to 1/2 case except for the
	${\cal B}_{bc}^{\prime}$ decays. Both the $c$ quark and $b$ quark decays of ${\cal B}_{bc}^{\prime}$ baryons are comparable to the corresponding 1/2 to 1/2 cases.
	\item ${\cal B}(H_{bc}\to H_{bs}l\nu)\sim10^{-2}$ holds for the corresponding
	results in Refs. \cite{Wang:2017mqp} and \cite{Shi:2016gqt}, while in this work, it ranges from $10^{-3}$
	to $10^{-2}$. ${\cal B}(H_{bc}\to H_{bd}l\nu)\sim10^{-3}$ holds
	for the corresponding results in Refs. \cite{Wang:2017mqp} and \cite{Shi:2016gqt}, while in this work,
	it ranges from $10^{-4}$ to $10^{-3}$. Here $H_{bc}$ stands
	for the $B_{c}$ meson or the ${\cal B}_{bc}^{(\prime)}$ baryon.
	\item ${\cal B}(H_{bc}\to H_{bs}\pi)\sim10^{-2}$ holds for the corresponding
	results in Refs. \cite{Wang:2017mqp} and \cite{Shi:2016gqt}, while in this work, it is roughly $10^{-3}$.
	${\cal B}(H_{bc}\to H_{bd}\pi)\sim10^{-3}$ holds for the corresponding
	results in Refs. \cite{Wang:2017mqp} and \cite{Shi:2016gqt}, while in this work, it is of order $10^{-4}$.
\end{itemize}

\subsection{Uncertainties}

We will also investigate the dependence of the decay widths on the
model parameters. Take $\Xi_{cc}^{++}\to\Sigma_{c}^{*+}$ transition
as an example. Varying the model parameters $m_{\{di\}}$, $\beta_{i}$,
$\beta_{f}$ and $m_{{\rm pole}}$ by 10\% respectively, the corresponding
error estimates are listed as follows
\begin{eqnarray}
{\cal B}(\Xi_{cc}^{++}\to\Sigma_{c}^{*+}e^{+}\nu_{e}) & = & (1.26\pm0.26\pm0.21\pm0.25\pm0.12)\times10^{-15}\ {\rm GeV},\nonumber \\
{\cal B}(\Xi_{cc}^{++}\to\Sigma_{c}^{*+}\pi^{+}) & = & (1.16\pm0.08\pm0.32\pm0.04\pm0.00)\times10^{-15}\ {\rm GeV},\nonumber \\
{\cal B}(\Xi_{cc}^{++}\to\Sigma_{c}^{*+}\rho^{+}) & = & (3.78\pm0.81\pm0.63\pm0.79\pm0.40)\times10^{-15}\ {\rm GeV},\nonumber \\
{\cal B}(\Xi_{cc}^{++}\to\Sigma_{c}^{*+}K^{+}) & = & (4.97\pm0.61\pm0.17\pm0.26\pm0.18)\times10^{-17}\ {\rm GeV},\nonumber \\
{\cal B}(\Xi_{cc}^{++}\to\Sigma_{c}^{*+}K^{*+}) & = & (1.60\pm0.44\pm0.59\pm0.49\pm0.25)\times10^{-16}\ {\rm GeV}.\label{eq:half_to_three_half}
\end{eqnarray}
For $\Xi_{cc}^{++}\to\Sigma_{c}^{+}$, the corresponding results are listed as follows
\begin{eqnarray}
{\cal B}(\Xi_{cc}^{++}\to\Sigma_{c}^{+}e^{+}\nu_{e}) & = & (1.04\pm0.04\pm0.02\pm0.11\pm0.15)\times10^{-14}\ {\rm GeV},\nonumber \\
{\cal B}(\Xi_{cc}^{++}\to\Sigma_{c}^{+}\pi^{+}) & = & (5.75\pm0.19\pm0.35\pm0.88\pm0.02)\times10^{-15}\ {\rm GeV},\nonumber \\
{\cal B}(\Xi_{cc}^{++}\to\Sigma_{c}^{+}\rho^{+}) & = & (2.61\pm0.08\pm0.08\pm0.29\pm0.27)\times10^{-14}\ {\rm GeV},\nonumber \\
{\cal B}(\Xi_{cc}^{++}\to\Sigma_{c}^{+}K^{+}) & = & (4.28\pm0.15\pm0.25\pm0.66\pm0.16)\times10^{-16}\ {\rm GeV},\nonumber \\
{\cal B}(\Xi_{cc}^{++}\to\Sigma_{c}^{+}K^{*+}) & = & (1.39\pm0.05\pm0.02\pm0.14\pm0.22)\times10^{-15}\ {\rm GeV}.\label{eq:half_to_half}
\end{eqnarray}
Some comments are given in order.
\begin{itemize}
	\item Eq. (\ref{eq:single_pole}) is also adopted for $\Xi_{cc}^{++}\to\Sigma_{c}^{+}$
	transitions for this time. In our previous work Ref. \cite{Wang:2017mqp}, the following
	fit formulas were adopted
	\begin{equation}
	F(q^{2})=\frac{F(0)}{1\mp\frac{q^{2}}{m_{{\rm fit}}^{2}}+\delta\left(\frac{q^{2}}{m_{{\rm fit}}^{2}}\right)^{2}}.\label{eq:fit_formula}
	\end{equation}
	However, only a few percent is changed in Eqs. (\ref{eq:half_to_half}) compared with our previous
	results in Ref. \cite{Wang:2017mqp}.
	\item It can be seen that, the variation in these parameters may cause a
	sizable change in the decay width, but the order of magnitude will
	not change.
\end{itemize}

\section{Conclusions}

In our previous work, we have  performed the calculation of doubly heavy
baryon weak decays for 1/2 to 1/2 case. As a continuation, we investigate
the 1/2 to 3/2 case in this work. Light-front approach under the diquark picture
is once again adopted to extract the form factors. In Ref.~\cite{Zhao:2018zcb},
the same method was used to study the bottom and charm baryon decays and reasonable results were obtained. The extracted form factors are then
applied to predict the decay widths of simi-leptonic and nonleptonic decays. We find that the result for 1/2 to 3/2 case is roughly one order
of magnitude smaller than the corresponding 1/2 to 1/2 case
except for the ${\cal B}_{bc}^{\prime}$ decays. For ${\cal B}_{bc}^{\prime}$ baryons, both the $c$ quark and $b$ quark
decays are comparable to the corresponding 1/2 to 1/2 cases. SU(3) symmetry and sources of SU(3) symmetry breaking
for semi-leptonic decays are discussed. The error estimates are also performed.

It should be noted that the decay branching ratio is proportional
to the lifetime of the initial  baryon. However, as we have pointed out in Ref.~\cite{Wang:2017mqp}, there exist large uncertainties in the lifetimes
of these doubly heavy baryons. Our future work will aim to fix this problem.

\section*{Acknowledgements}

The author is grateful to Prof. Wei Wang for valuable discussions
and constant encourage- ments. This work is supported in part by National
Natural Science Foundation of China under Grant Nos. 11575110, 11655002,
11735010, Natural Science Foundation of Shanghai under Grant No. 15DZ2272100.

\end{document}